\documentclass{aastex631}

\begin{document}

\title{How the CME on 21 April 2023 Triggered the First Severe Geomagnetic Storm of Solar Cycle 25}

\correspondingauthor{Evangelos Paouris}
\email{Evangelos.Paouris@jhuapl.edu}

\author[0000-0002-8387-5202]{Evangelos Paouris}
\affiliation{The Johns Hopkins University Applied Physics Laboratory, 11100 Johns Hopkins Rd, Laurel, MD 20723, USA}

\author[0000-0002-8164-5948]{Angelos Vourlidas}
\affiliation{The Johns Hopkins University Applied Physics Laboratory, 11100 Johns Hopkins Rd, Laurel, MD 20723, USA}

\author[0000-0001-6913-1330]{Manolis K. Georgoulis}
\affiliation{The Johns Hopkins University Applied Physics Laboratory, 11100 Johns Hopkins Rd, Laurel, MD 20723, USA}

\author[0000-0002-8164-5948]{Phillip Hess}
\affiliation{US Naval Research Laboratory, Washington, DC 20375, USA}

\author[0000-0001-8480-947X]{Guillermo Stenborg}
\affiliation{The Johns Hopkins University Applied Physics Laboratory, 11100 Johns Hopkins Rd, Laurel, MD 20723, USA}

\begin{abstract}
The first severe (G4) geomagnetic storm of Solar Cycle 25 occurred on 23-24 April 2023, following the arrival of a Coronal Mass Ejection (CME) on 23 April. The characteristics of this CME, measured from coronagraphs (speed and mass), did not indicate that it would trigger such an intense geomagnetic storm. In this work, our aim is to understand why this CME led to such a geoeffective outcome. Our analysis spans from the source active region to the corona and inner heliosphere through 1 au using multiwavelength, multi-viewpoint remote sensing observations and in situ data. We find that rotation and possibly deflection of the CME resulted in an axial magnetic field nearly parallel to the ecliptic plane during the Earth encounter, which might explain the storm's severity. Additionally, we find that imaging away from the Sun-Earth line is crucial in hindcasting the CME Time-of-Arrival at Earth. The position (0.39 au) and detailed images from the SoloHI telescope onboard the Solar Orbiter mission, in combination with SOHO and STEREO images, helped decisively with the three-dimensional (3D) reconstruction of the CME.
\end{abstract}


\section{Introduction} \label{sec:Section_1_Introduction}

As the current Solar Cycle (SC25) approaches its peak and solar eruptive activity intensifies, so do the space weather effects of this activity. The first severe geomagnetic storm of SC25 occurred on 23-24 April 2023, reaching a minimum Dst of -213 nT and a Kp index of 8+. This was the first time that the Dst index reached below -200 nT since the 17 March 2015 geomagnetic storm \citep[][]{Wu_2016_StPatricks_Storm, Marubashi_2016_StPatricks_Storm}. The current fleet of spacecraft distributed in the inner heliosphere, equipped with in situ and remote sensing instruments, allow us to study potentially geoeffective events with higher fidelity than in the past.

In this work, we demonstrate these new capabilities by analyzing the April 2023 storm. It was triggered by the arrival of a Coronal Mass Ejection (CME) first registered in coronagraphic images at around 18:00 UT on 21 April 2023. The rather typical coronal characteristics of the event (i.e. speed and mass) did not indicate that it could be capable of triggering a severe G4 geomagnetic storm. From a space weather perspective, this event is unusual. The typical expectation is that severe geomagnetic storms result from CMEs associated with X-class solar flares and fast halo CMEs, like the recent events of May 2024 \citep[see e.g.][]{Liu_2024_May2024_Superstorm, Kontogiannis_2024_May2024_Superstorm, Jarolim_2024_May2024_Superstorm}.

The April 2023 CME was recently studied by \cite{Weiss_2024_April2023Storm} who applied a novel distorted magnetic flux rope (DMFR) model to reconstruct the complex geometry of this CME, as observed in situ by the Wind and STEREO-A spacecraft. Their results demonstrate significant deviations from conventional cylindrical or toroidal approximations, emphasizing the utility of their model for capturing MFR geometry. However, while \cite{Weiss_2024_April2023Storm} study primarily focuses on the magnetic structure and validation of the model, our study is concerned with the space weather impact and geoeffectiveness of the event.

To understand why this event turned out to be intensely geoeffective, we explore the conditions that influenced the development of the active region, examine the characteristics of the unstable magnetic flux rope (MFR) and the resulting CME in the low corona and near the Sun, and finally, analyze the inner heliospheric signatures obtained from in situ measurements at multiple locations. A CME is defined as an eruption of a coherent, magnetic coronal structure \citep[][]{Vourlidas_et_al_2013_CMEs_Fluxropes}, where the magnetic field lines are twisted around a single common axis. The region at the center of this structure is usually designated as the magnetic flux rope of the CME \citep[MFR; see e.g.][and references therein]{Weiss_2024_April2023Storm}. In addition to observations from coronagraphs and heliospheric imagers, it can be tracked with in situ observations, where it is commonly referred to as the magnetic obstacle \citep[MO;][]{Nieves-Chinchilla_2018_Wind_MCs} or magnetic ejecta \citep[ME;][]{Lugaz_2012_deflection_ME}. When part of the MFR exhibits clear signatures of a rotating magnetic field component combined with low plasma beta values, it is referred to as a magnetic cloud \citep[MC;][]{Burlaga_1981_MC, Bothmer_&_Schwenn_1998}.


The motivation behind this work is two-fold: First, we want to understand why this CME, with rather average characteristics measured by coronagraphs, led to a severe geomagnetic storm. This understanding is crucial for improving our ability to assess the potential geoeffectiveness of CMEs on the basis of their initial conditions. Second, we seek to evaluate and improve the accuracy of CME Time-of-Arrival (ToA) predictions by performing a hindcast using multi-viewpoint remote sensing data. By doing so, we demonstrate the importance of incorporating observations from multiple vantage points, such as those provided by Solar Orbiter's SoloHI instrument, in enhancing the reliability of space weather forecasting models for Earth-directed CMEs where projection effects can hinder accurate assessment.

For the first objective, we study the magnetic complexity of the source solar active region. We then use remote sensing observations to determine the MFR's chirality and the toroidal field's inclination and direction. Subsequently, we verify the MFR type using in situ measurements at the Sun-Earth Lagrangian point L1. For the second objective, we track the CME in space via a 3D reconstruction for as long as possible and then perform a retrospective forecast of its ToA at Earth.

The paper is organized as follows. Section 2 describes the data and instruments utilized, both remote sensing and in situ. In Section 3, we present an analysis of the active region and the MFR orientation at the Sun and at 1 au utilizing multiple data sets. Section 4 focuses on the analysis of the CME (3D reconstruction, kinematics, and energetics) and the retrospective forecast of its ToA. In Section 5, we discuss our findings and present our conclusions.

\section{Data and Observations} \label{sec:Section_2_Data_and_Observations}

We use white-light data from i) the Large Angle and Spectroscopic Coronagraph \citep[LASCO;][]{Brueckneretal1995} onboard the Solar and Heliospheric Observatory \citep[SOHO;][]{Domingoetal1995}, ii) the Sun-Earth Connection Coronal and Heliospheric Investigation (SECCHI) instruments \citep{Howard_SECCHI_2008} onboard the Solar TErrestrial RElations Observatory (STEREO) mission \citep{Kaiseretal2008}, and iii) the Heliospheric Imager \citep[SoloHI;][]{Howard_etal_2020_SolOHI} of the Solar Orbiter mission \citep[SolO;][]{Muller_etal_2020_Solar_Orbiter}. For a detailed comparison of SoloHI images with LASCO and SECCHI, see Table 1 in \citet{Hess_2023} and the surrounding text.

\begin{figure*}[ht]
\includegraphics[width=0.95\textwidth]{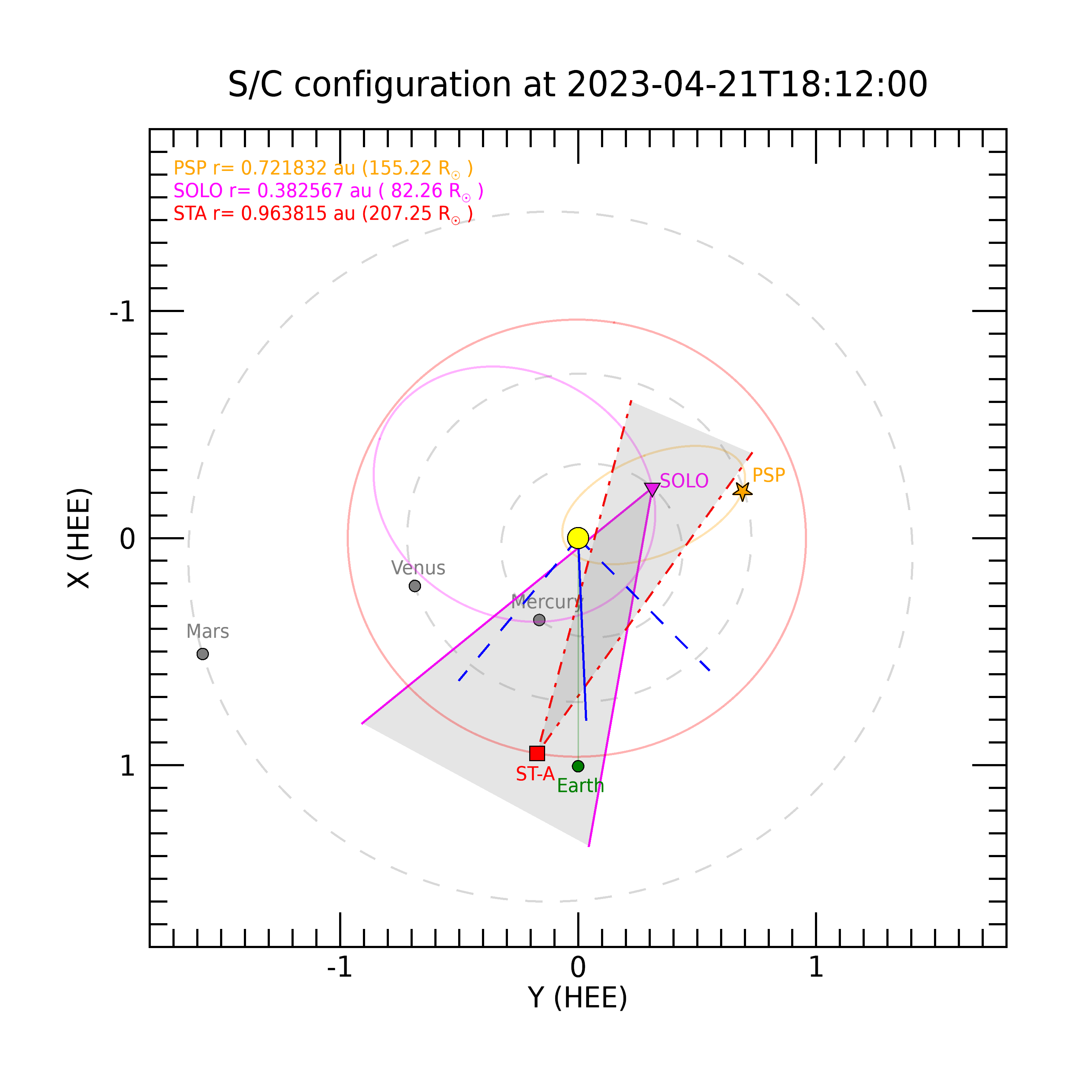}
\centering
\caption{Spacecraft configuration on April 23, 2023, at the time of the CME onset (18:12 UT) projected on Heliocentric Earth Ecliptic (HEE) coordinates. The solid blue line indicates the propagation direction of the CME, while the dashed blue lines show its angular width. The SoloHI field of view (FOV) is represented by a gray triangle enclosed by solid magenta lines, and the STEREO-A HI1 FOV is between the red dash-dot lines. Both heliospheric imagers captured the CME from nearly 180$\circ$ apart, but its propagation direction could not be ascertained robustly by either instrument alone. The combination of the observations, however, provided much better results and was critical for our analysis. The Parker Solar Probe was near its aphelion, so no WISPR observations are available.} 
\label{fig:SC_orientation}
\end{figure*}

The April 21, 2023, CME was observed by all available coronagraphs including LASCO C2-C3/SOHO and COR2/STEREO. Additionally, the CME was captured by the two heliospheric imagers (HI) onboard STEREO, HI1 and HI2, and the HI onboard SolO nearly $180^{\circ}$ apart. The spacecraft configuration is presented in Figure~\ref{fig:SC_orientation}.  

Multi-wavelength EUV images and line-of-sight magnetograms were taken from the Atmospheric Imaging Assembly \citep[AIA;][]{Lemenetal2012_AIA} and the Helioseismic and Magnetic Imager \citep[HMI;][]{Scherreretal2012}, both onboard the Solar Dynamics Observatory mission \citep[SDO;][]{Pesnelletal2012}. We also use EUV images from the Extreme UltraViolet Imager \citep[EUVI;][]{Wuelser_STA_EUVI_2004, Howard_SECCHI_2008} onboard STEREO.

The in-situ measurements are taken from the Wind spacecraft. In particular, we used data from the Magnetic Fields Investigation \citep[MFI;][]{Lepping_1995_MFI_Wind}, the Solar Wind Experiment \citep[SWE;][]{Ogilvie_1995_SWE_Wind}, and the 3-D Plasma and Energetic Particle Investigation \citep[3DP;][]{Lin_1995_PM_3DP_Wind} instruments. STEREO-A in-situ data are taken from the In-situ Measurements of PArticles and CME Transients/Magnetometer\citep[IMPACT/MAG;][]{Luhmann_2008_IMPACT} and the Magnetic Field Vector and PLAsma and SupraThermal Ion Composition \citep[PLASTIC;][]{Galvin_2008_PLASTIC}.

\section{Active Region Evolution and MFR Type Analysis}
\label{sec:Section_3_AR_and_Flux_Rope_Analysis}
\subsection{Evolution of the Active Region} \label{subsec:Subsection_3_1_AR_analysis}
NOAA active region 13283 (AR13283), which produced the M1.7-class solar flare associated with the CME of interest, initially appeared on the eastern limb of the Sun on April 14, 2023, following the complex NOAA AR13281. It remained visible in the southern hemisphere until April 26. AR13283 is responsible for the first severe geomagnetic storm of Solar Cycle 25, so it is important to investigate its magnetic evolution and complexity characteristics. To better understand the evolution of AR13283, we calculated the effective connected magnetic field strength, $B_{eff}$, at the solar photosphere \citep{GeorgoulisandRust2007}.

\begin{figure*}[ht]
\includegraphics[width=0.45\textwidth]{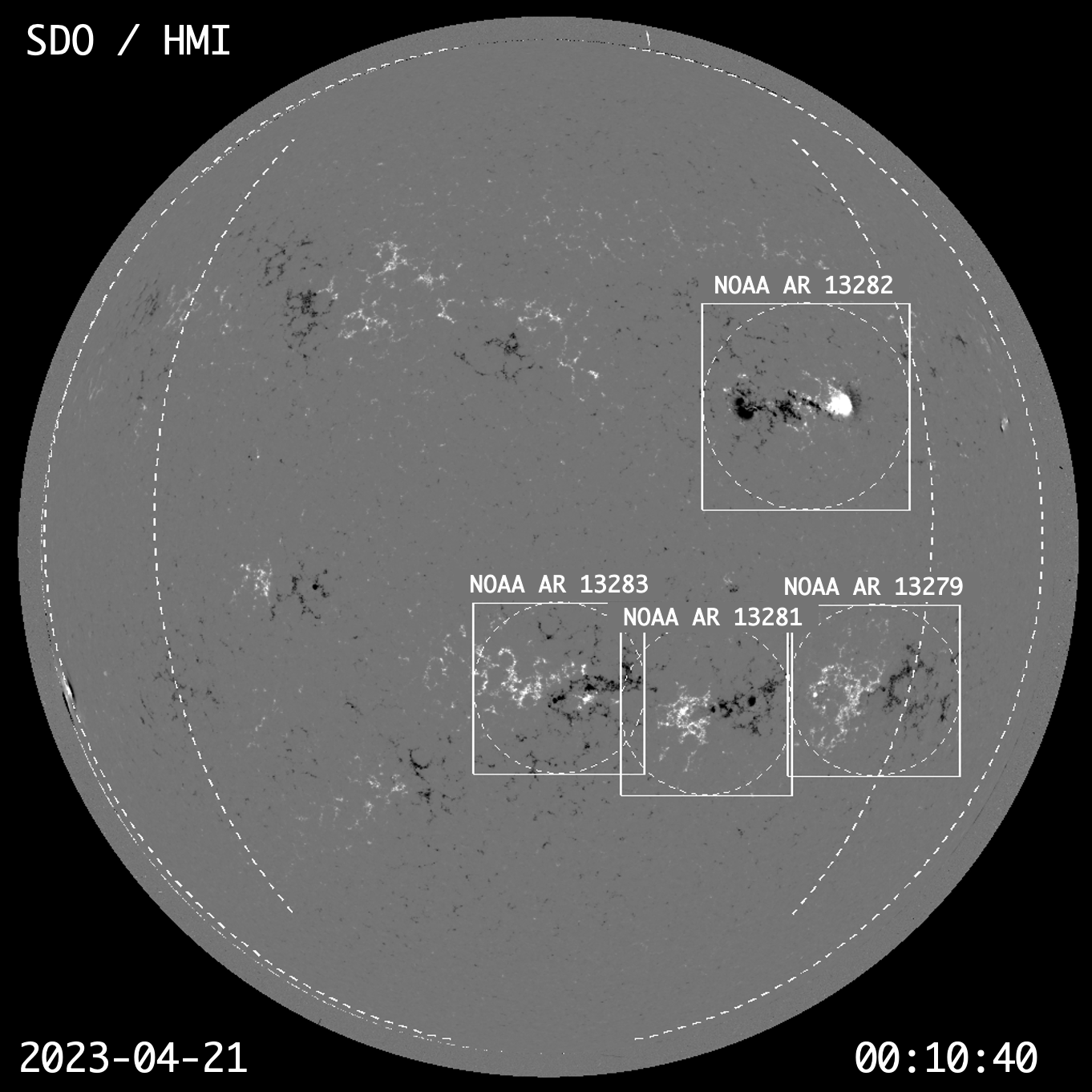}
\includegraphics[width=0.45\textwidth]{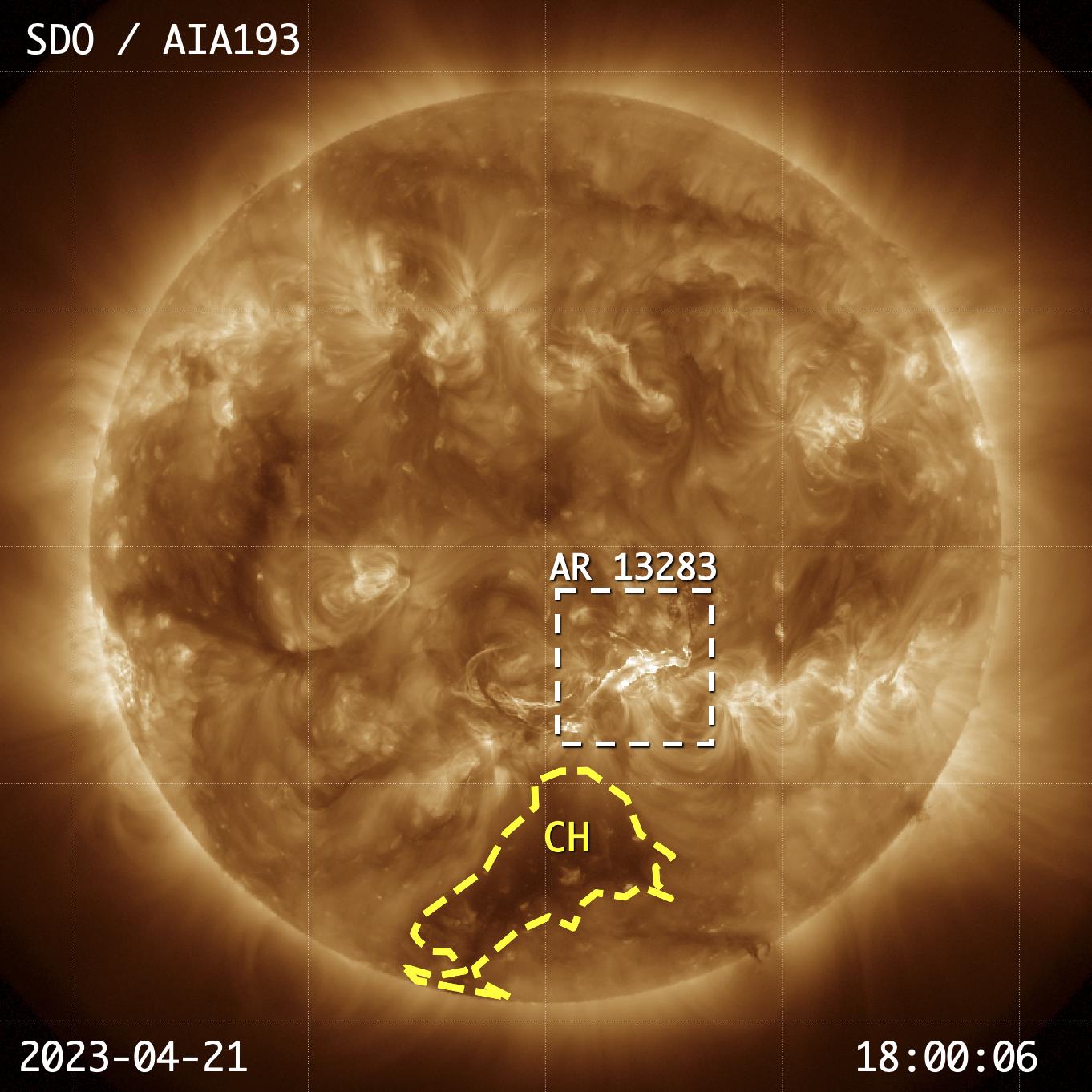}
\includegraphics[width=0.5\textwidth]{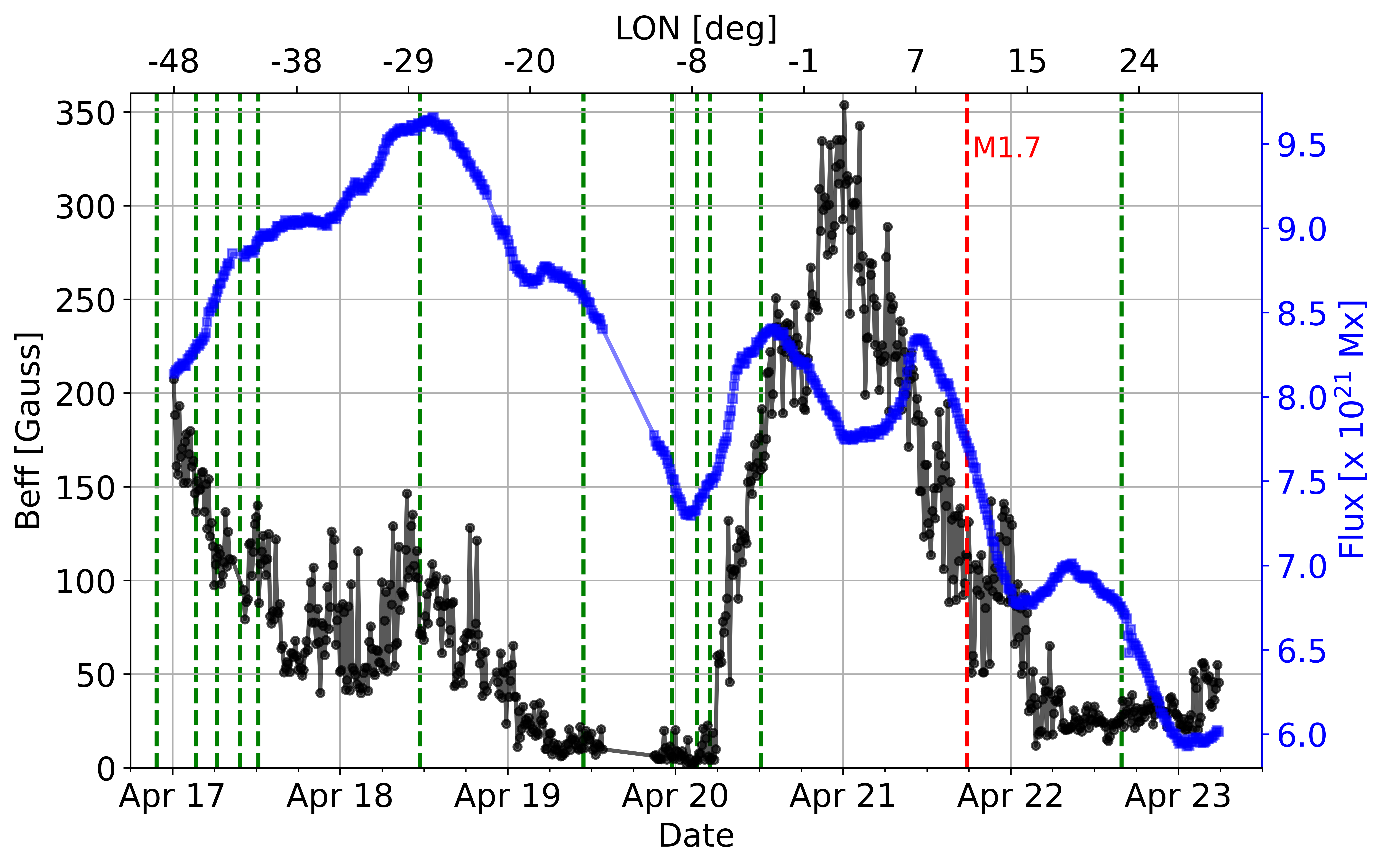}
\centering
\caption{Top left panel: ARIA output applied to a full-disk solar magnetogram on April 21, 2023, at 00:10 UT. Four active regions fulfilled our selection criteria, namely NOAA ARs 13279, 13281, 13282, and 13283. Dashed circumscribed circles and attached squares indicate each of these regions. The dashed areas in the eastern and western hemispheres indicate central meridian distances of $\pm60^\circ$ (inner) and 
$\pm80^\circ$ (outer).  
Top right panel: A full-disk AIA 193~\AA~image taken close to the peak time of the M1.7 solar flare. The position of AR~13283 is marked with a white dashed-line square. The coronal hole, located south of AR~13283 and responsible for the deflection of the CME (as discussed later), is highlighted with a yellow dashed line.
Low middle panel: The calculated $B_{eff}$ values and the total unsigned magnetic flux with a temporal resolution of 12 minutes, as a function of time (bottom abscissa) and longitude (top abscissa), in AR~13283. The vertical-colored dashed lines indicate the start times of the solar flares triggered in the region, with green for C-class and red for M-class.} 
\label{fig:ARIA_and_Beff}
\end{figure*}

We processed magnetograms from the HMI via the Active Region Identification Algorithm \citep[ARIA; see e.g.,][]{Georgoulis_etal_2008_ARIA} to track the region of interest (ROI) and calculate its $B_{eff}$. In Figure~\ref{fig:ARIA_and_Beff}, the left panel displays a result of the ARIA algorithm for a given time, when a maximum $B_{eff}=350$ G was calculated. The right panel illustrates $B_{eff}$ and the central heliographic longitude of AR13283 as a function of time. The time evolution of the total unsigned magnetic flux is shown in blue. We observed an increase in the flux profile from April 17 to mid-April 18 and again on April 20, indicating flux emergence. In both cases, the flux emergence coincides with the occurrence of 70\% of the total observed C-class flares. However, flux cancellation, which coincides with a decrease in $B_{eff}$ observed around mid-April 21, suggests unstable conditions in the region where the magnetic MFR was located just before the eruption. The flaring activity of the active region for C-class (M-class) solar flares is indicated by the vertical green (red) dashed lines. The $B_{eff}$ value of AR13283 increased rapidly from almost 10 G on April 20, 2023, at 06:00 UT, to approximately 350 G on April 21, 2023, at 00:30 UT. The M1.7-class solar flare occurred on April 21, 2023, at 18:12 UT (peak time), at heliographic coordinates S21W12.

Given the significant geomagnetic impact, one might initially expect that the source active region was highly complex expressed through high $B_{eff}$ values. \cite{Georgoulis2008} have shown that M-class flares tend to originate from ARs with substantial magnetic complexity, where the peak preflare $B_{eff}$ values for M-class solar flares ranged between 630~G and 2640~G. Similarly, \cite{Paourisetal_5Sep2022} analyzed AR13088, which produced a fast and massive CME, and found a maximum $B_{eff}$ of approximately 2500~G. The high magnetic complexity in that region was associated with abundant flare activity, including more than 17 M-class and 50 C-class flares, further supporting the idea that high $B_{eff}$ values correlate with frequent and powerful eruptions.

However, the case of AR13283 presents an intriguing counterexample. Despite reaching a maximum $B_{eff}$ value of only 350~G, the CME produced a severe geomagnetic storm with a minimum Dst of -213 nT.
In fact, this AR produced only 15 solar flares, the majority of which were C-class, with a single M-class flare (M1.7). The relatively low magnetic complexity, as indicated by the modest $B_{eff}$, and flaring activity, suggests that the region was less magnetically complex than the ARs studied by \cite{Georgoulis2008} and \cite{Paourisetal_5Sep2022}, raising the question of how such a severe geomagnetic storm could arise from an apparently ``mediocre" active region.

It is worth noting that the active region calculation pipeline for $B_{eff}$ in \cite{Georgoulis2008} utilized magnetograms from the Michelson Doppler Imager \citep[MDI;][]{Scherrer_1995_MDISOHO} onboard SOHO. As we used magnetograms from HMI/SDO, some differences in the algorithm outputs are expected to calculate $B_{eff}$. However, these differences are insignificant in terms of magnitude, so the maximum calculated $B_{eff}$ can still be used to interpret the complexity of AR by comparing it with the results of \cite{Georgoulis2008}. 
\cite{Liu_2012_HMI_vs_MDI} demonstrated that the line-of-sight magnetic signal derived from the MDI data is stronger than that from the HMI by a factor of 1.40. 
To investigate how this discrepancy affects our calculations, we performed the following exercise: we selected three different intervals on 20 April 2023 when the $B_{eff}$-values were low (01:58 - 03:46 UT), moderate (11:22 - 13:10 UT) and high (20:58-22:46 UT), as per Figure \ref{fig:ARIA_and_Beff}. We chose 10 HMI magnetograms of AR13283 out of each of these intervals and produced two additional sets of them: one in which each of the 30 HMI magnetograms was emulated to look like an MDI magnetogram, but at HMI resolution (“MDI-hr” in Table \ref{tab:Beff_HMI_vs_MDI}) and another in which each of the 30 MDI-emulated magnetograms was reduced to the nominal MDI spatial resolution of 3.96 arcsec (1.98 arcsec/pixel --  “MDI-lr” in Table \ref{tab:Beff_HMI_vs_MDI}). Table 1 shows the means of the $B_{eff}$-values and respective standard deviations (in parentheses).  
Our findings support the hypothesis that the HMI and MDI $B_{eff}$ values are comparable, although there is an expected trend for increased $B_{eff}$ values as we go from HMI to MDI-emulated magnetograms at HMI resolution to nominal MDI-resolution magnetograms, in line with the findings of \cite{Liu_2012_HMI_vs_MDI} for field strength.

\begin{deluxetable*}{cccc}[ht]
\tablecaption{Mean $B_{eff}$ values (with standard deviations in parentheses) for different input magnetograms at different time intervals (see text for explanations). Means and standard deviations occur from a sample of 10 magnetograms for each category.}
\label{tab:Beff_HMI_vs_MDI}
\tablewidth{0pt}
\tablehead{
\colhead{Input} & \colhead{Low} & \colhead{Moderate} & \colhead{High} 
}
\startdata
HMI    & 47.6 (4.6)   & 131.9 (13.1) & 165.8 (16.6) \\
MDI-hr & 53.2 (10.6)  & 119.7 (15.8) & 179.2 (48.6) \\
MDI-lr & 66.0 (11.7)  & 132.4 (8.6)  & 205.0 (42.4) \\
\enddata 
\end{deluxetable*}

The extracted portions of ARIA serve as input for calculating $B_{eff}$. We observed that about 18 hours before the eruption, $B_{eff}$ reached its maximum value of approximately 350 G. 

\subsection{The magnetic flux rope orientation at the Sun} \label{subsec:Subsection_3_2_Flux_Rope_I}

Here we characterize the MFR of the CME by identifying the following parameters: chirality, axis orientation, and axial field direction. 

\begin{figure*}[ht]
\includegraphics[width=1.0\textwidth]{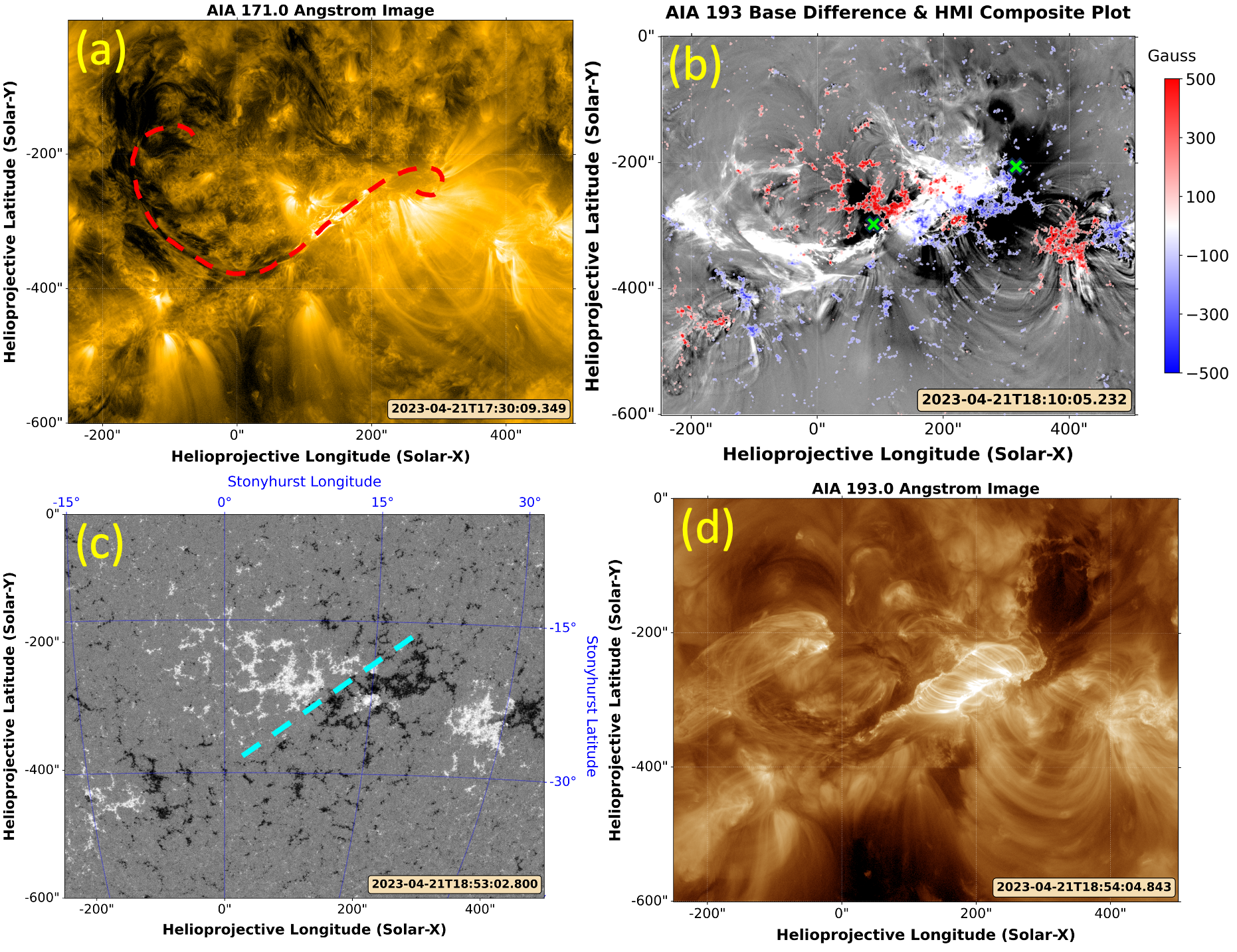}
\centering
\caption{(a) AIA image in 171~\AA~showing the forward-S filament shape (red dashed curve) indicating right-handed chirality. (b) Base difference 193~\AA~AIA image with overlaid HMI magnetogram contours (blue/red for negative/positive polarity). The post-eruption dimming regions (signatures of the MFR footpoints) are indicated with green crosses. The difference has been taken between the images at 17:09:59~UT and 18:09:59~UT on 21 April. (c) HMI magnetogram centered on AR13283, with the main polarity inversion line (PIL) approximated by the cyan dashed line.  
(d) AIA 193~\AA~image showing that the post-eruption arcades (PEAs) are oriented similarly to the PIL. The inclination of the MFR axis with respect to the ecliptic is $|\tau| \sim 34^{\circ}$, indicating a low-inclination MFR.
} 
\label{fig:AIA_HMI_Multi}
\end{figure*}

We carefully analyzed the source active region and the evolution of the eruptive filament to estimate the MFR chirality. The eruptive filament is visible on SDO/AIA 131, 171, 193, and 304 Å channel images. It is mainly located east of AR13283 in the southern solar hemisphere. A forward-S filament shape is observed in the AIA 171 Å channel, indicating right-handed chirality (see Figure~\ref{fig:AIA_HMI_Multi}a). After the eruption, the filament quickly regrouped and remained above the active region for more than a week until it finally erupted on April 28 around 11:00~UT as shown, e.g., in the composite wavelet-processed \citep[][]{Stenborg_2008_Wavelets} 195 and 304 EUVI frames from STEREO-A available at \url{https://solar.jhuapl.edu/Data-Products/EUVI-Wavelets.php}. The first brightening, signaling the beginning of the eruption, began around 17:27~UT. 
An EUV wave is visible on AIA 171 Å and 193 Å images at 17:53~UT. In Figure~\ref{fig:AIA_HMI_Multi}b we present a base-difference 193 Å image of the region of interest overlaid on the HMI magnetogram contours (blue/red is used for negative/positive polarity). The dimming regions, the signatures of the MFR footpoints \citep[][]{CME_footpoints_Thompson_2000}, are depicted with green crosses. In Figure~\ref{fig:AIA_HMI_Multi}c we show the HMI magnetogram and the polarity inversion line (PIL), which is approximated by a cyan dashed straight line. The post-eruption arcade (PEA) loops are visible and align with the PIL, following the filament direction above the AR (see Figure~\ref{fig:AIA_HMI_Multi}d). The post-eruption dimming is visible for more than an hour on AIA 193 Å images, starting at 18:15~UT, just after the peak of the M1.7 class solar flare.  

To determine the initial alignment of the MFR, we consider the positions of the PIL and the PEAs. The MFR axis is considered almost parallel to the PIL \citep[][]{Marubashi_2015_PIL_and_PEAs} or the PEA \citep[][]{Yurchyshyn_2008_PEAs_and_flux_rope}. Following the eruption of an MFR, a reconnection process occurs, leading to the formation of flare ribbons over the footprints of the PEAs observed in EUV \citep[][]{Tripathi_2004_EUV_PEAs}. Since both the PIL and the PEA were clearly visible, we define the MFR inclination with respect to the ecliptic as $\tau$, the average of the orientations from both the PIL and PEA: $|\tau| = (|\tau_{\text{PIL}}| + |\tau_{\text{PEA}}|) / 2$ \citep[][]{Palmerio_etal_2017}. The tilt angle $\tau$ is measured from the solar East and assumes a positive (negative) value if the acute angle to the ecliptic is to the north (south). The PIL is approximated as a straight line, and we estimate $|\tau_{\text{PIL}}| \sim 35^{\circ}$. The inclination of the PEAs is $|\tau_{\text{PEA}}| \sim 33^{\circ}$. Given the small difference, we estimate that the uncertainty should be less than $5^{\circ}$ \citep[][]{Palmerio_etal_2018} and the inclination of the MFR is $|\tau| \sim 34^{\circ}$. In both cases (PIL and PEAs), the inclination measurements were taken when the AR was located very close to the central meridian, minimizing the projection effects.

For a major geomagnetic impact, a strong long-duration negative $B_{z}$ component is a necessary condition. This is more likely when the central axis of the MFR is approximately parallel to the ecliptic plane (low-inclination MFRs) since the $B_{z}$ component in such cases represents the helical field, and its sign changes as the CME crosses the observer \citep[see e.g.][]{Bothmer_&_Schwenn_1998}. It can thus maintain a southward orientation for an extended period. However, high-inclination MFRs, with their central axis at large angles to the ecliptic, can also produce strong geomagnetic storms if the axial field is directed southward. For the event under discussion, the $|\tau| \sim 34^{\circ}$ places the MFR in the low-inclination category, making it a candidate for a strong geomagnetic impact.

Finally, we carefully inspected the regular, running, and base-difference AIA 193 A images to associate the coronal dimmings with their respective magnetic polarities. We found that the lower eastern dimming (see Figure~\ref{fig:AIA_HMI_Multi}b,d) is rooted in positive magnetic polarity, while the dimming on the west is rooted in negative magnetic polarity. Thus, the axial field is directed from the positive footprint to the negative one, or from left to right as we observe the AR.

Based on these parameters (chirality, axis orientation, and direction of the axial field) we assess that the erupting  MFR type is a right-handed MFR (RH) south-west-north (SWN) MFR \citep{Bothmer_&_Schwenn_1998}:.

\subsection{Magnetic flux rope identification in situ} \label{subsec:Subsection_3_3_Flux_Rope_II}

\begin{figure*}[ht]
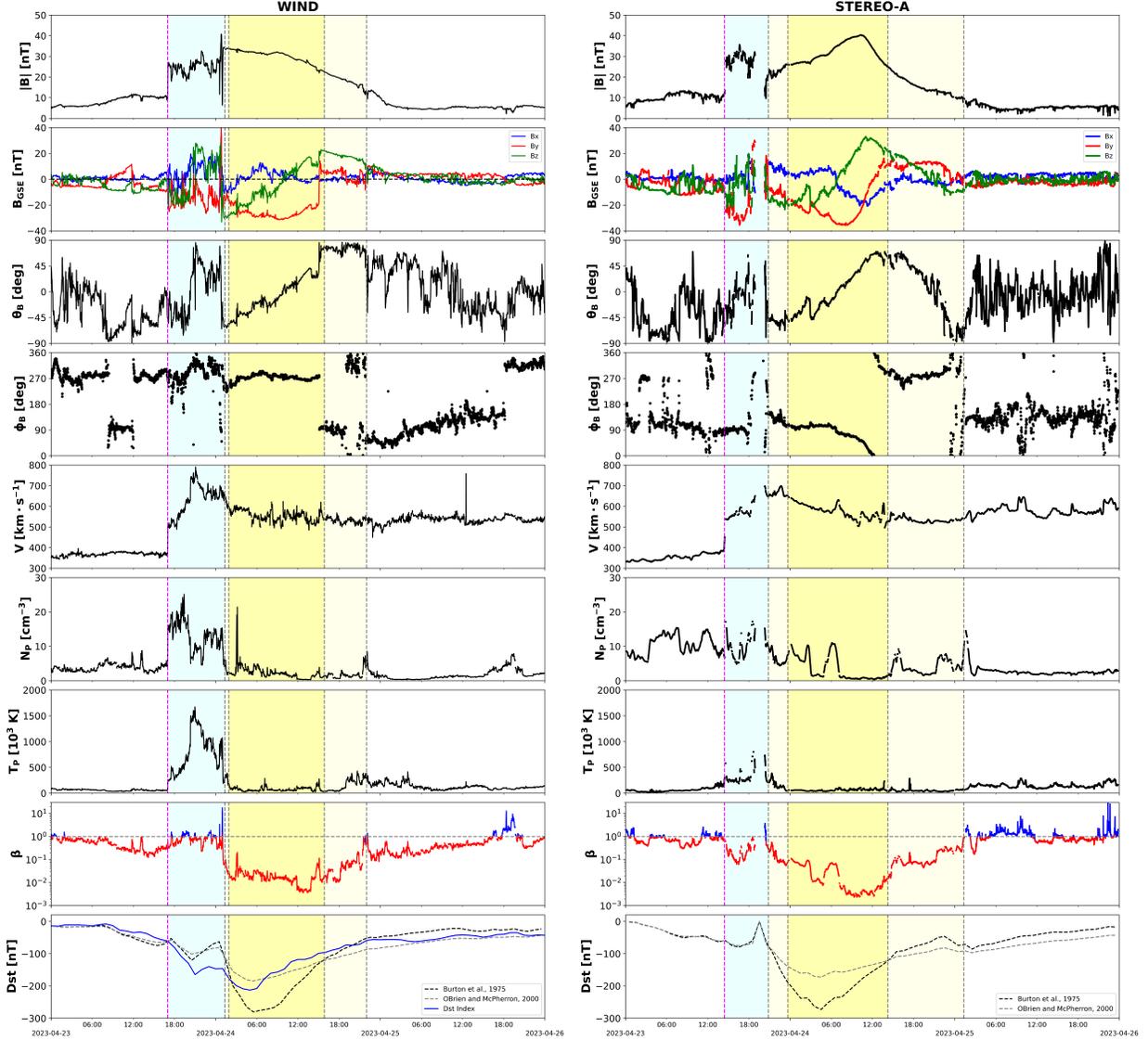

\includegraphics[width=0.45\textwidth]{Fig4a.png}
\includegraphics[width=0.45\textwidth]{Fig4b.png}
\centering
\caption{In situ measurements of solar wind magnetic fields and plasma parameters from Wind (left) and STEREO (right) spacecraft. The parameters shown from top to bottom are the total magnetic field magnitude, the magnetic field components ($B_{x}$, $B_{y}$, and $B_{z}$) in GSE coordinates, $B_{\theta}$ and $B_{\phi}$ components in GSE angular coordinates for Wind and the transformed components from RTN for STEREO (see text for details). Then, we show the solar wind speed, proton density, and plasma $\beta$. The last panel for both spacecraft shows the calculated Dst index values using the \cite{Burton_1975_Dst} (black dashed line) and \cite{OBrien_McPherron_2000_Dst} (gray dashed line) models. The observed Dst index values (blue line) are presented on the plot on the left (Wind). The vertical magenta dashed line indicates the interplanetary shock, while the vertical gray dashed lines indicate the times associated with the boundaries of the ME and the MC. The sheath region is highlighted in light blue, the ME region is highlighted in light yellow (including the MC part), and the MC is highlighted in yellow.} 
\label{fig:In_situ_Wind_STEREO}
\end{figure*}

\begin{figure*}[ht]
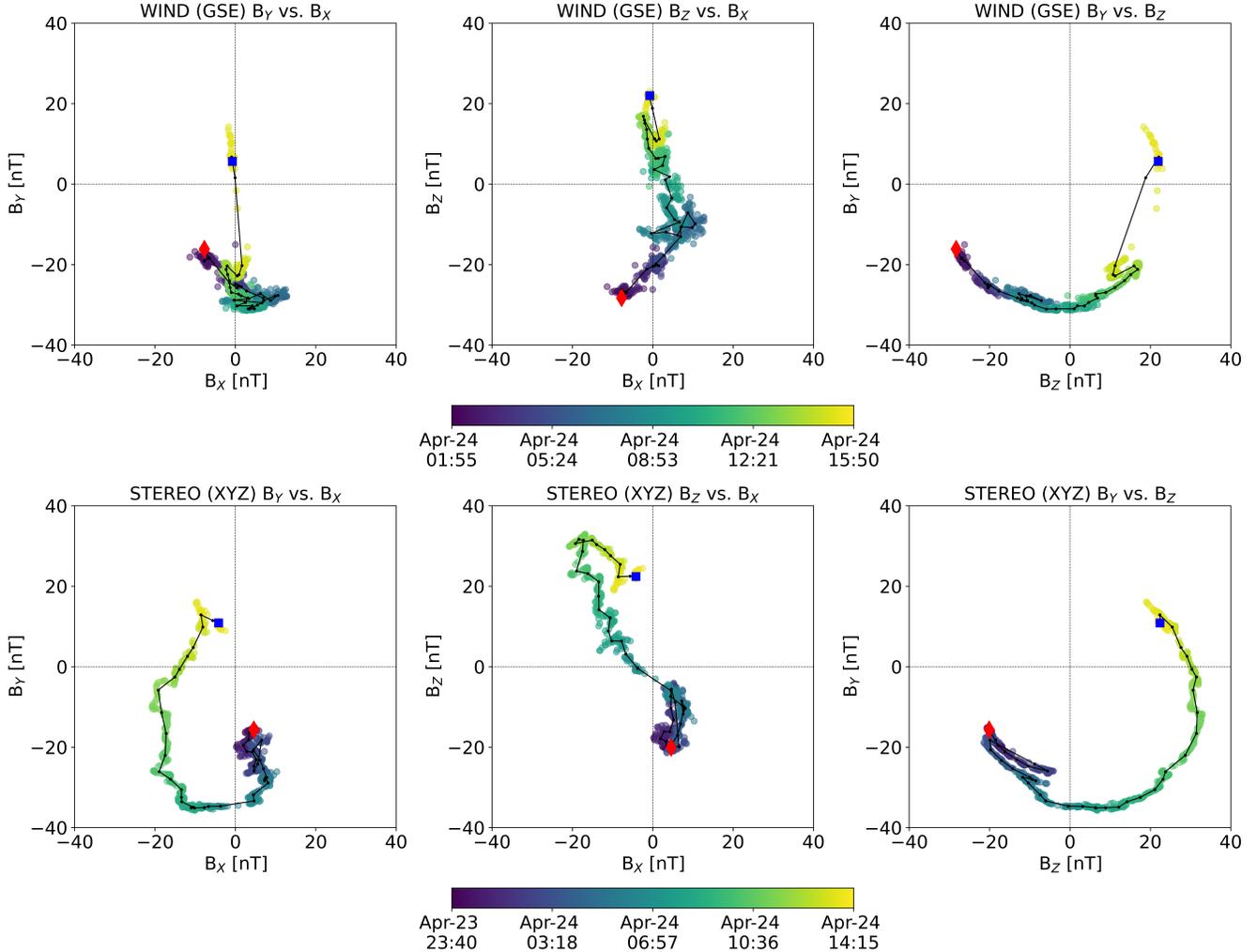

\includegraphics[scale=0.4]{Fig5a.png}
\includegraphics[scale=0.4]{Fig5b.png}
\caption{Magnetic hodograms of the GSE magnetic field components for the MFR part of the CME at L1, using Wind (top row) and STEREO (bottom row) in situ data. The 20-minute average values are overplotted with a black line. A red diamond indicates the beginning of the rotation, and a blue square indicates the end of the rotation.} 
\label{fig:Hodograms_comb}
\end{figure*}

\subsubsection{Magnetic flux rope signatures at Wind} \label{subsec:Subsection_3_3_1_Flux_Rope_II_Wind}

The next step is to analyze the CME magnetic field components at the Wind/L1 position, leveraging STEREO-A observations as an additional comparison point within the inner heliosphere. The interplanetary shock associated with the CME was detected by Wind on 23 April 2023 at 17:00 UT, as indicated by the vertical dashed magenta line in the left panel of Figure~\ref{fig:In_situ_Wind_STEREO}. The ME signatures were identified on 24 April, spanning approximately 01:20 UT to 22:00 UT, with MC signatures, characterized by the rotation of at least one magnetic field component and a low plasma $\beta$, observed from approximately 01:55 UT to 15:50 UT on April 24. In the left panel of Figure~\ref{fig:In_situ_Wind_STEREO}, the sheath region is highlighted in light blue, the ME region in light yellow, and the MC in yellow.

We use the Cartesian ($B_{x}$, $B_{y}$, $B_{z}$) and angular ($\theta_{B}$, $\phi_{B}$) magnetic field data in the geocentric solar ecliptic (GSE) coordinate system. A visual examination of the data (see Figure~\ref{fig:In_situ_Wind_STEREO} left panel) indicates that the MFR characteristics at L1 are similar to the coronal ones; i.e., the $B_{z}$ component is initially negative (south), then smoothly rotates and finally becomes positive (north). The $B_{y}$ component is consistently negative (west) during this rotation, while the $B_{x}$ component does not show a significant rotation. The rotations of the magnetic field components can be seen in the magnetic hodograms of Figure~\ref{fig:Hodograms_comb} \citep[see e.g.][]{Nieves-Chinchilla_2018_Wind_MCs}. 
The Figure shows combined hodograms of pairs of GSE magnetic field components. Overplotted with the black line is the 20-minute average data; the start time is marked with a red diamond, and the end time is marked with a blue square. The third dimension (color scale) helps to track the change of the magnetic field components as a function of time. All plots correspond to the MFR interval. Following the black line, the MFR configuration is confirmed by an almost $180^{\circ}$ rotation of the magnetic field vector in the z–y plane (see $B_{y}$–$B_{z}$ hodogram).

To estimate the orientation of the MFR axis and its helicity sign, we apply the minimum variance analysis \citep[MVA;][]{Sonnerup_Cahill_1967_MVA} to the 20-min averaged in situ measurements on the observed MC where the magnetic field rotations are clear. This technique has been used in many cases on in situ data \citep[see e.g.][]{Palmerio_etal_2017}, and despite its caveats and constraints \citep[see e.g.][]{Dunlop_1995_MVA, Huttunen_2005_MCs}, it is considered reliable. From MVA, we estimate the orientation of the MFR axis at 1 au by determining the latitude $\theta_{MVA}$ and longitude $\phi_{MVA}$ in angular coordinates. The MFR axis corresponds to the intermediate variance direction, where $\theta_{MVA}=90^{\circ}$ is defined as northward and $\phi_{MVA}=90^{\circ}$ is defined as eastward. Thus, the latitude $\theta_{MVA}$ can be used to estimate the inclination of the axis with respect to the ecliptic plane. The helicity sign is inferred from the rotation of the magnetic field components as the MFR passes over the spacecraft. Specifically, a right-handed MFR will exhibit a specific rotation pattern in hodograms, such as a counterclockwise rotation when moving from negative to positive $B_{z}$ in the $B_{y}$–$B_{z}$ plane (see Figure~\ref{fig:Hodograms_comb} third panel, first row). The findings from the application of the MVA analysis are considered consistent by checking that the ratio of intermediate to minimum eigenvalues, $\lambda_2/\lambda_3$, is greater than 2 \citep[see e.g.][]{Bothmer_&_Schwenn_1998, Huttunen_2005_MCs}. Finally, we find that the orientation of the MFR axis obtained from the MVA to be ($\theta_{MVA}$, $\phi_{MVA}$) = ($-16^{\circ}$, $109^{\circ}$). It is consistent with a low inclined MFR. The ratio $\lambda_2/\lambda_3$ is approximately 5.3, confirming the method's validity. The MVA fittings assume an uncertainty of $10^{\circ}$ as the toroidal magnetic field of the CME crosses the spacecraft almost perpendicularly \citep[][]{Palmerio_etal_2018}.

The final step in our analysis is to estimate the distance of the spacecraft from the central axis and the apex of the MFR. The former is based on the ratio of the minimum-variance direction to the total magnetic field in the MVA frame \citep[][]{Demoulin_Dasso_2009_MCs}, i.e., $\langle|B_{min}|\rangle / \langle B \rangle$, where the averages are considered over the MFR interval. A higher ratio indicates a progressively distant crossing from the MFR central axis, signaling that the inferred MFR orientation may deviate from its true value \citep[][]{Palmerio_etal_2018}. In our case, we found a ratio of approximately 0.20, indicating that the spacecraft crossed the MFR relatively closer to the central axis. The latter is based on the proxy of the location angle \citep[$\lambda$;][]{Janvier_2013_MCs}, where $\sin \lambda = \cos \theta_{MVA} \cdot \cos \phi_{MVA}$. The location angle $\lambda$ ranges from $-90^{\circ}$ (close to the first leg of the CME) to $90^{\circ}$ (close to the second leg of the CME). If $\lambda = 0^{\circ}$, the spacecraft crossed the CME near the center of the MFR. We found $\lambda \sim -19^{\circ}$, indicating that the spacecraft crossed the CME relatively close to the center.

As mentioned in the Introduction, during the severe geomagnetic storm, the Dst index reached a minimum of $-213$~nT and the Kp index reached 8+. The observed minimum Dst coincides with the interval of the MC, as expected (see Figure~\ref{fig:In_situ_Wind_STEREO}, last panel on the left figure). The orientation of the MFR indicates a low-incline structure, and the $B_{z}$ component was expected to change sign as the CME crossed Earth, assuming that the MFR maintained a consistent orientation throughout its transit. We also used Wind in situ data to estimate the Dst index values using widely applied models of \citep{Burton_1975_Dst} and \cite{OBrien_McPherron_2000_Dst}. The \citep{Burton_1975_Dst} model tends to overestimate the actual Dst index, predicting a minimum Dst of -281 nT, while the \cite{OBrien_McPherron_2000_Dst} model underestimates it, with a minimum Dst of -185 nT. The latter is closer to the observed minimum value of -213 nT. The combination of both models provides a reliable bounding range for the observed values of the Dst index.

\subsubsection{Magnetic flux rope signatures at STEREO} \label{subsec:Subsection_3_3_2_Flux_Rope_II_STEREO}

The propagation direction of the CME was favorable for in situ detection by STEREO-A. The interplanetary shock was detected by STEREO-A on April 23, 2023, at 14:25 UT, 2.6 hours before it arrived at Wind. The MC signatures were identified from approximately 23:40 UT on April 23 to 14:15 UT on April 24. The spacecraft were at similar distances from the Sun, around 0.964~au for STEREO and 0.997~au for Wind. The small longitudinal separation between Wind and STEREO ($\sim 10^{\circ}$) led to some differences in the two  in situ profiles (Figure~\ref{fig:In_situ_Wind_STEREO}).

The STEREO-A data were transformed to the GSE frame for comparison to the Wind profiles. This transformation does not affect the $B_{z}$ component. Generally, the transformation $B_{R} = -B_{x}$, $B_{T} = -B_{y}$, and $B_{N} = B_{z}$ can be applied to approximate the GSE coordinates. Although this is not an exact GSE transformation due to the non-terrestrial reference frame, it provides a comparable coordinate system, but it is valid only for locations close to the ecliptic plane. This approximation is applicable in this case because Wind and STEREO-A were close to the ecliptic plane, at heliospheric latitudes of $-4.9^{\circ}$ and $-5.8^{\circ}$, respectively.

We used the MC interval for the STEREO-A MVA analysis, as we did with Wind, and found that the orientation of the MFR axis was ($\theta_{MVA}$, $\phi_{MVA}$) = ($-17^{\circ}$, $71^{\circ}$), consistent with a low-inclination MFR. The ratio $\lambda_2/\lambda_3 \approx 18$, confirming the validity of the method. The STEREO-A crossing distance was very close to the MFR central axis given the low ratio of the minimum variance direction to the total magnetic field in the MVA frame (0.09). The location angle of $\lambda \sim 18^{\circ}$, suggests that STEREO-A crossed near the MFR apex towards the eastern leg of the MFR in contrast to the Wind results. Thus, the Wind and STEREO-A MFR crossings differ by almost $40^{\circ}$, which may partially explain the variations in the in situ profiles. Overall, the Wind and STEREO-A MVA results are in agreement indicating a low inclination MFR and supporting our hypothesis that the strength of the storm is the result of the MFR orientation.

The differences between the Wind and STEREO-A data provide interesting information. The $\sim 10^{\circ}$ angular longitudinal separation translates to a distance of approximately 0.17~au. The 2.6-hour difference between the shock arrival at STEREO-A and then Wind, and the 0.033~au radial separation indicates a shock speed of 530 km/s, consistent with the in situ solar wind speed observed by both spacecraft. 
The sheath region was longer in Wind (8.3 hours) compared to STEREO-A (6.4 hours) but the ejecta region was longer by almost 40\% in STEREO-A (28.5 hours) than in Wind (20.7 hours). The duration of the MFR was nearly identical (13.9 hours at Wind and 14.6 hours at STEREO). In the Wind data, we observe rotation in the magnetic field component $B_{z}$, with typical SWN flux rope signatures. In STEREO, both $B_{y}$ and $B_{z}$ components show a sign change, with an SWN flux rope signature present during the MC interval. However, the magnetic field components and the total magnetic field magnitude, $|B|$, in STEREO exhibit a more complex structure that differs significantly from what is observed in Wind.

We estimated the Dst index at STEREO-A using the \cite{Burton_1975_Dst} and \cite{OBrien_McPherron_2000_Dst} models. The primary motivation for this approach is to assess the potential "geoeffectiveness" of the CME impact at the STEREO-A location, which is $10^\circ$ east from the Sun-Earth line. We found a minimum Dst at STEREO-A of -273 nT using the \cite{Burton_1975_Dst} model and -173 nT using the \cite{OBrien_McPherron_2000_Dst} model, highlighting a potential similarity in impact between STEREO-A and Earth (see Figure~\ref{fig:In_situ_Wind_STEREO}, bottom right panel). This finding is valuable for space weather forecasting, as despite the differences observed in situ measurements between Wind and STEREO-A, our results indicate that the overall impact at both locations could be comparable.

\section{CME Analysis: Kinematics, 3D Reconstruction and Retrospective Arrival Forecast}
\label{sec:Section_4_CME_Analysis}
\subsection{Kinematics and Energetics}
\label{subsec:Subsection_4_1_Kinematics}

\begin{figure*}[ht]
\includegraphics[width=0.75\textwidth]{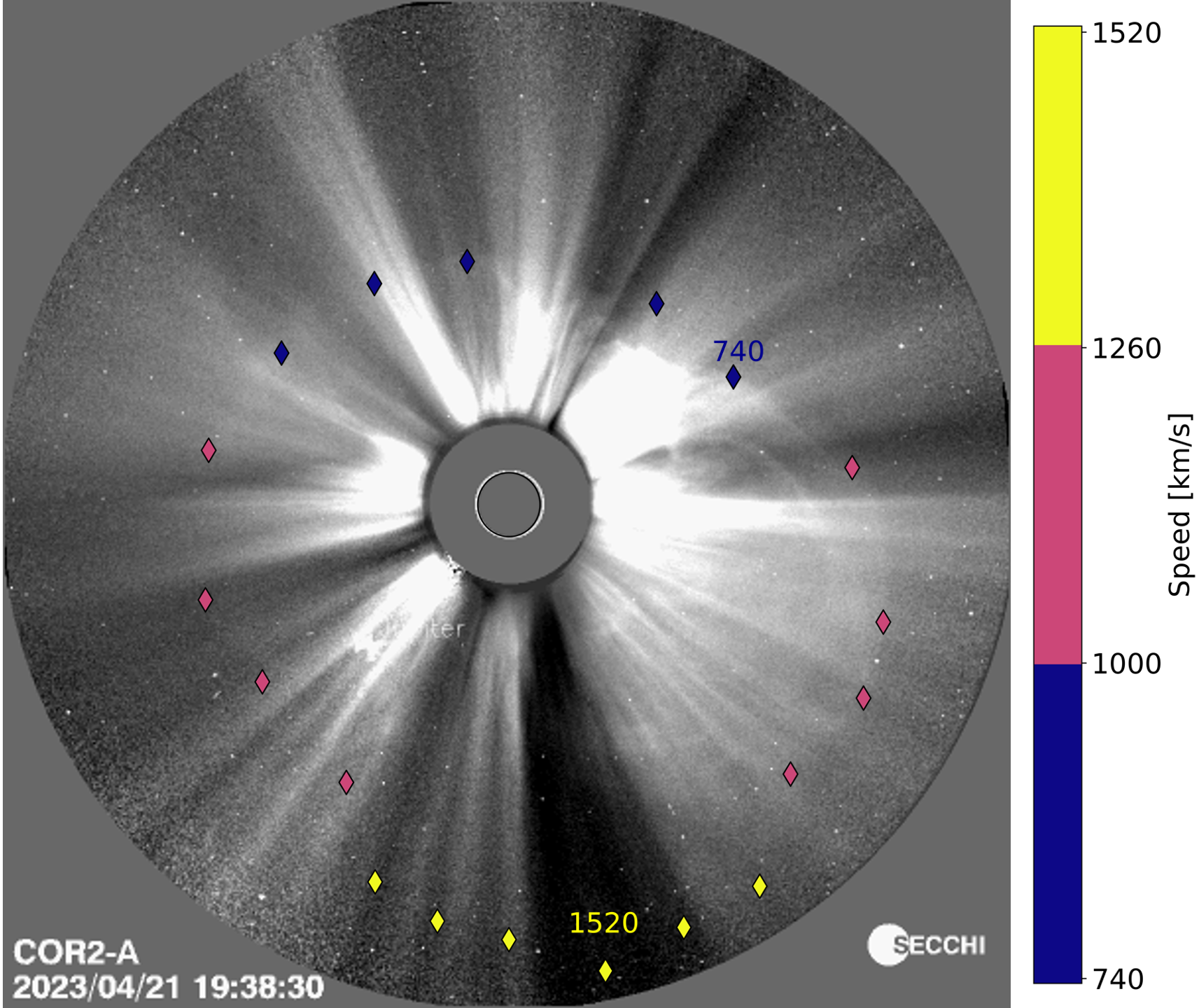}
\centering
\caption{Distribution of the estimated speeds across the CME front on top of a COR2 frame at 19:38 UT. The CME is expanding asymmetrically. The highest speeds (above 1200 km/s) are observed along the southern front of the CME shock with a maximum speed of 1520 km/s. The lowest speeds are observed along the northern front, with a minimum speed of 740 km/s.}
\label{fig:COR2_Kinematics}
\end{figure*}

The CME was observed from both the STEREO and SOHO spacecraft as a halo. It appeared in the C2 and COR2 images on April 23, 2023, at 18:12 UT and 18:23 UT, respectively. We used COR2 data to estimate the plane-of-sky (POS) speed of the CME, at multiple position angles (PAs), covering the entire angular extent ($0^{\circ}$ to $360^{\circ}$). This is the optimal approach for estimating kinematics for halo CMEs \citep[][]{Paourisetal_5Sep2022}, since halo CMEs are most affected by projection effects \citep[see e.g.,][and references therein]{Paouris_2021_ProjectionEffects}. Our single-viewpoint analysis indicated an asymmetric expansion, with the CME moving faster towards the south, where the highest speeds, depicted in yellow color in Figure~\ref{fig:COR2_Kinematics}, yield an average value of 1400 km/s with a standard deviation of 90 km/s. In contrast, towards the north, the lowest speeds, shown in blue in Figure~\ref{fig:COR2_Kinematics}, correspond to an average value of 820 km/s with a standard deviation of 80 km/s. Closer to the ecliptic plane, the CME speed was approximately 1000 km/s, indicating a moderately fast CME (see Figure~\ref{fig:COR2_Kinematics}). However, these speeds do not suggest a strong geoeffective response. They are consistent with the in-situ measurements at 1~au, where the CME shock arrival at Wind on April 23, 2023, at 17:00~UT corresponds to a transit time of $\sim47$ hours and an average transit speed of 880 km/s. We also used the COR2 observations to estimate the CME mass and kinetic energy budget \citep[see][for details of the procedure]{Vourlidas_etal_2010_CME_mass_energy}.

We track the CME up to $15~R_\odot$, when the outer front, associated with the shock of the CME, leaves the COR2 FOV. The maximum mass is 7.4 $\times$ 10$^{15}$ g. Using the maximum calculated speed of 1520 km/s, we obtain a kinetic energy estimate of 8.6 $\times$ 10$^{31}$ erg. Note that these values represent lower estimates due to projection effects \citep{Vourlidas_etal_2010_CME_mass_energy}. 

\subsection{3D Reconstrunction - GCS} 
\label{subsec:Subsection_4_2_GCS}

The CME kinematic measurements in the previous section revealed a range of projected speeds.  To derive a deprojected speed of the CME, we turn to 3D reconstruction of the CME flux-rope structure to estimate its direction of propagation and the height of the apex as a function of time. Following standard practice, we apply the Graduated Cylindrical Shell model \citep[GCS;][]{Thernisien_2006_GCS, Thernisien_2011_GCS} on observations between April 21, 18:20 UT and April 22, 06:30 UT using all available white light imagers (i.e., LASCO/SOHO, SECCHI/STEREO, and SoloHI/SolO).

\begin{figure*}[ht]
\includegraphics[width=0.95\textwidth]{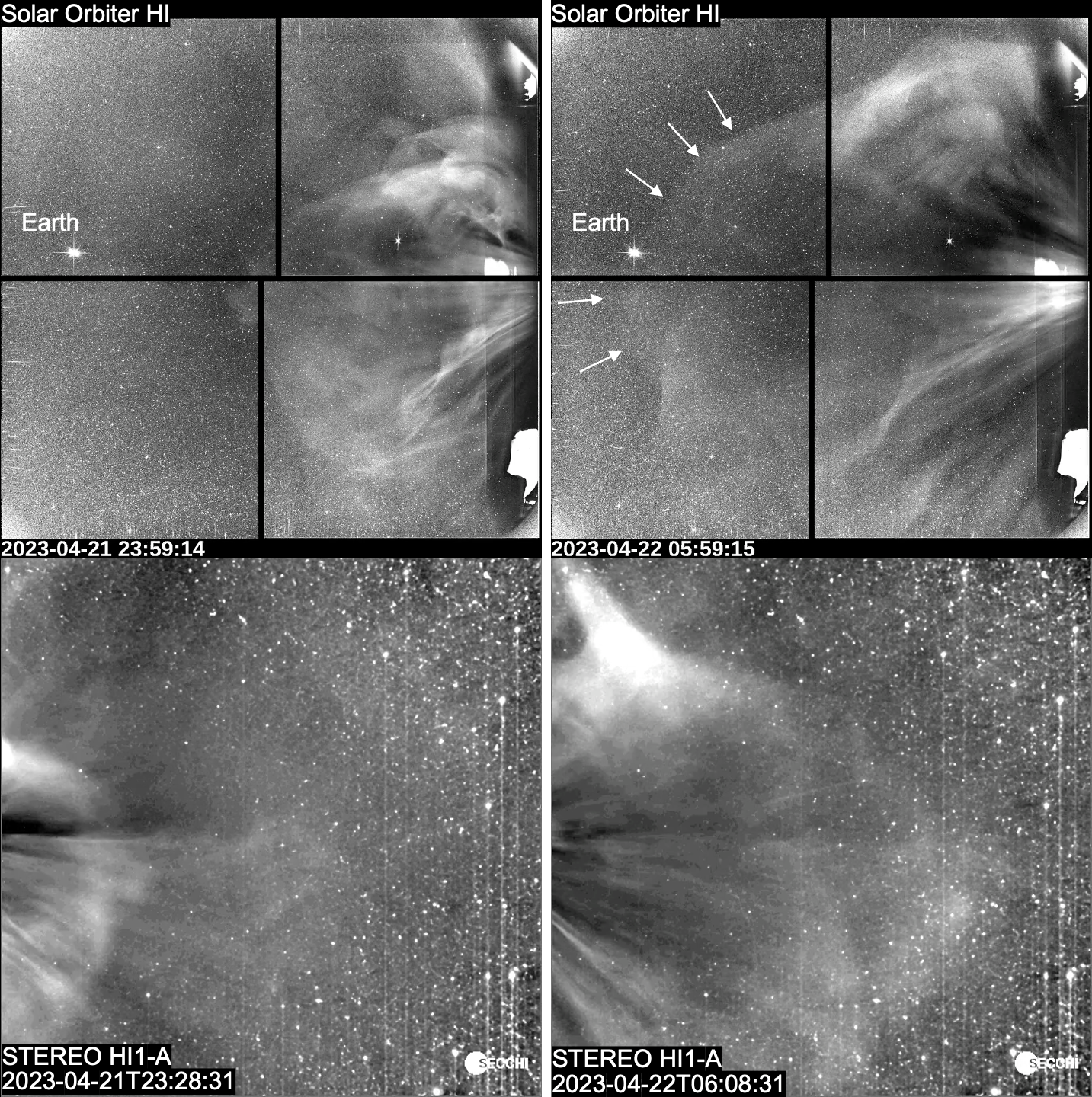}
\centering
\caption{Top row: SoloHI images of the CME on April 21 at 23:59 UT (left panel) and April 22 at 05:59 UT (right panel). They show the shock and CME substructures in detail. Earth is visible as the bright spot in the upper left tile. The white arrows in the right panel indicate the shock front. Bottom row: STEREO-A Heliospheric Imager-1 (HI1) images at times closest to the SoloHI images. The leading edge of the CME is seen in both instruments early on (left column). However, the CME fronts becomes diffuse in HI1 later on (right panel). It remains clear in SoloHI enabling better tracking and improved arrival forecast as we discuss in the text.
}
\label{fig:SoloHI_STEREOHI_detail_image}
\end{figure*}

Despite the availability of observations, it was not straightforward to assess the potential impact of the CME. Because this was an Earth (and STEREO-A) directed event, projection effects hindered the accurate determination of the event kinematics using images from these missions alone. Fortunately, the SolO spacecraft was positioned off the Sun-Earth line, allowing SoloHI to capture the CME in detail (see the images in the top row of Figure~\ref{fig:SoloHI_STEREOHI_detail_image}). Tracking the CME into the inner heliosphere from HI1 was challenging, as the CME boundaries became diffuse in the HI1 FOV (see the bottom row of Figure~\ref{fig:SoloHI_STEREOHI_detail_image}), in contrast to the clearer images from SoloHI. The overall CME shape is similar in the two instrument images because they observe the CME from opposite sides (Figure~\ref{fig:SC_orientation}). The SoloHI viewing perspective and high-resolution imaging provide the information necessary to robustly reconstruct the 3D CME trajectory with the inclusion of the LASCO and SECCHI imaging.

\begin{figure*}[ht]
\includegraphics[width=0.95\textwidth]{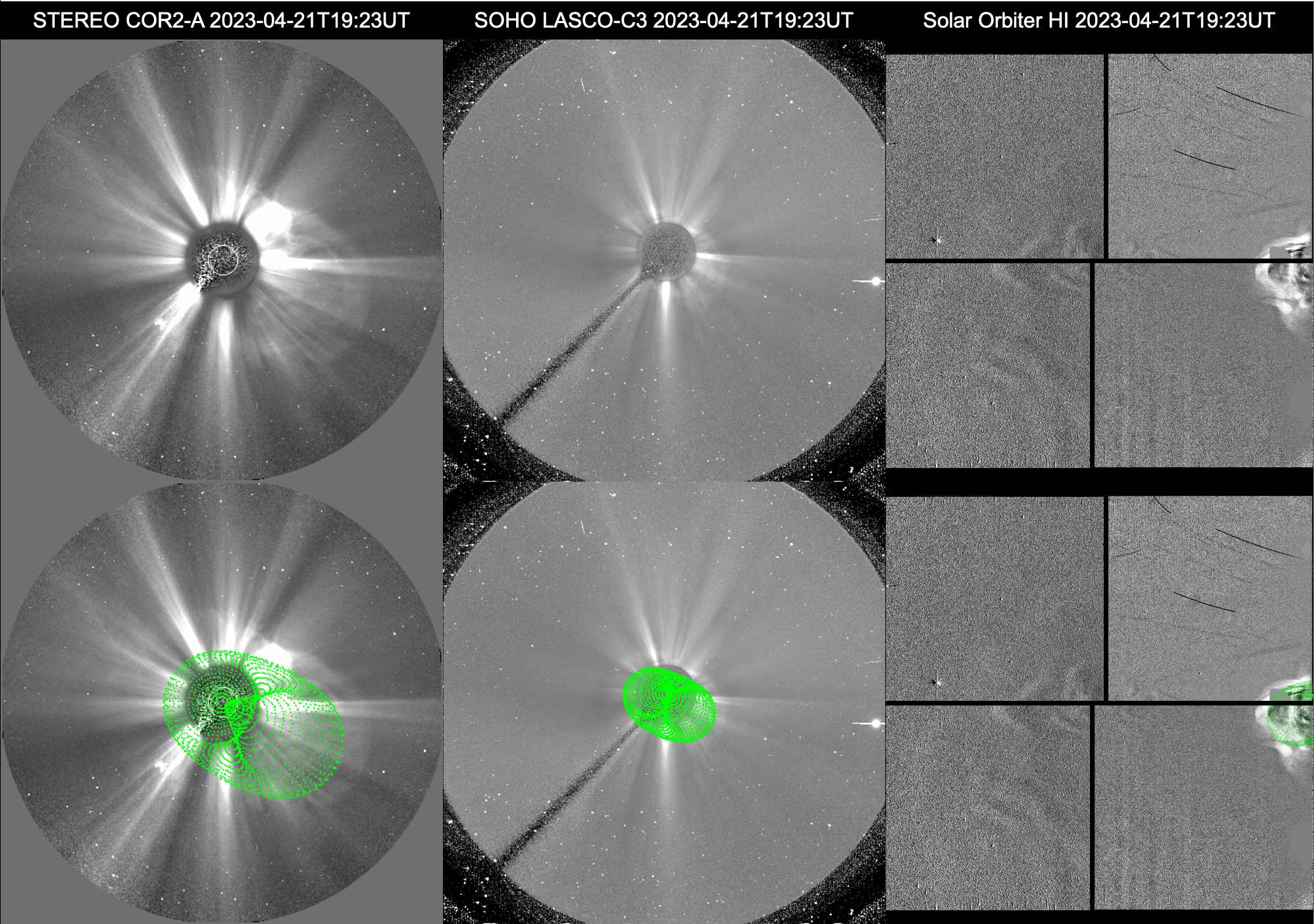}
\centering
\caption{CME observations on 2023 April 21 around 19:23 (upper row) and GCS mesh grid (green dots) projected over each observation (bottom row). Images and GCS projections are for STEREO COR2-A (left column), SOHO LASCO C3 (middle column), and Solar Orbiter HI (right column).}
\label{fig:GCS_COR2_C3_SoloHI_image}
\end{figure*}

Based on the GCS reconstruction, we estimate that the CME propagated along a Carrington longitude of $350^{\circ}$ and a latitude of $-13^{\circ}$. For reference, Earth was at a Carrington longitude of roughly $345^{\circ}$ at the onset of the event. The CME tilt is $-20^{\circ}$ with a half-angular width\footnote{From the GCS modeling, the CME exhibited an aspect ratio, $\kappa$, of approximately 0.37 and a half-angle, $\alpha$, of approximately $19^{\circ}$ \citep[for a detailed description of the GCS parameters, see][]{Thernisien_2011_GCS}.} of $41^{\circ}$. 
We use images for the GCS reconstruction from three different viewpoints, i.e., SOHO at Lagrange L1, STEREO-A at a distance 0.96~au positioned $\sim~10^{\circ}$ to the east of the Sun-Earth line, and SolO at a distance 0.38~au positioned $\sim~125^{\circ}$ west of the Sun-Earth line (see Figure~\ref{fig:SC_orientation}). A snapshot of the 3D reconstruction on 2023 April 21, 19:23~UT is shown in Figure~\ref{fig:GCS_COR2_C3_SoloHI_image}. The use of three viewpoints for the GCS reconstruction implies uncertainties of about $\sim~2^{\circ}$ for latitude, $\sim~4^{\circ}$ for longitude, while the uncertainties are larger for tilt and angular width ($2.6^{\circ}-22^{\circ}$ and $4.7^{\circ}-10.6^{\circ}$, respectively) \citep[see e.g.][]{Kay_2024_LLAMACORE, Verbeke_2023, Thernisien_2009_GCS_and_Errors}. The height of the apex ranged from $5.8~R_\odot$ (April 21, 18:23 UT) to $85.3~R_\odot$ (April 22, 06:34 UT).

\subsection{Retrospective forecast of the arrival of the CME} 

The SoloHI viewing perspective and high-resolution data were crucial for tracking the CME up to approximately $85~R_\odot$. This is critical for applying the two-phase kinematics approach \citep[HeRPA;][]{Paouris_Vourlidas_2022}. Many Time-of-Arrival (ToA) forecasting algorithms rely on simple empirical relations to represent interplanetary propagation, primarily using kinematic information from coronagraphic observations below $30~R_\odot$ and making several simplifying assumptions about constant direction and speed for the transient. In \cite{Paouris_Vourlidas_2022}, the assumption of constant speed in the inner heliosphere was replaced with a realistic scenario of a two-phase behavior, consisting of a decelerating (or accelerating) phase from the first available point in heliospheric imagers to a certain distance (i.e., the last available point), followed by a coasting phase to Earth (or Mars).

We applied the two-phase kinematics approach to the deprojected height measurements of the CME based on the GCS output. This facilitates a kinematic analysis by applying a second-degree polynomial fit to the data, resulting in an observed deceleration of approximately -21 m/s². Assuming that the CME progresses at a nearly constant speed beyond the last deprojected height of $85~R_\odot$, we can use this kinematic analysis to make a hindcast prediction of the CME ToA. Our estimate, which includes a deceleration phase followed by coasting to 1 AU, predicts a ToA on April 23, 2023, at approximately 15:13 UT. This is only 1.8 hours earlier than the actual ToA of the CME measured at Wind. 

In contrast, a hindcast estimate based on GCS fit using only LASCO and SECCHI predicted a ToA on April 24, 2023, at approximately 08:44~UT, fully 15.7 hours later than observed. 

\begin{figure*}[ht]
\includegraphics[width=0.95\textwidth]{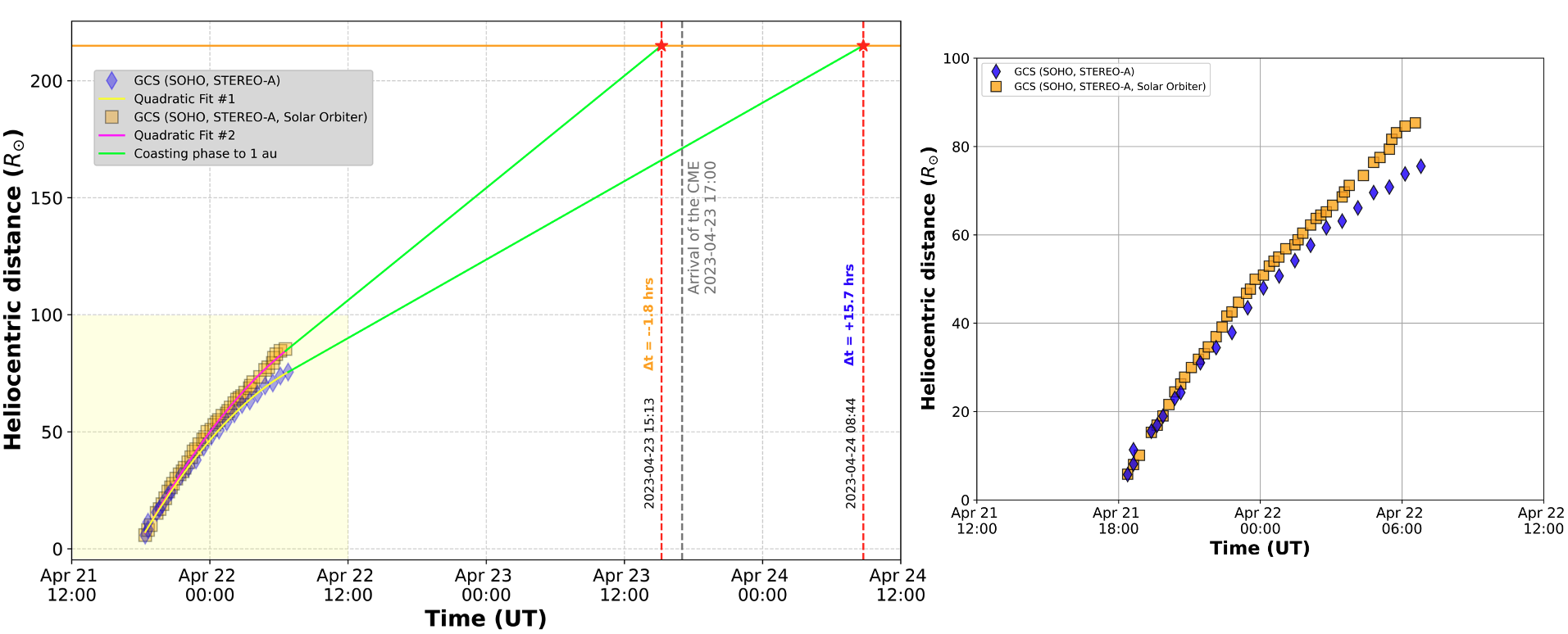}
\centering
\caption{
Two-phase kinematics of the April 21, 2023, CME based on 3D reconstructions from the GCS using two viewpoints (SOHO and STEREO; blue diamonds) and three viewpoints including SolO (orange squares). The left panel displays the complete kinematic profile, with the deceleration phase represented by a second-degree polynomial fit (yellow and magenta lines, respectively) to the deprojected heliocentric distances (square and diamond points) as a function of time. A coasting phase (lime lines) is assumed from the last available data point up to 1 AU. The region of interest, where the points are densely clustered and difficult to distinguish, is highlighted in yellow in the left panel. This region is shown in greater detail in the right panel, which provides a zoomed-in view to better visualize the differences between the 3D reconstructions with and without SolO data. Our initial estimation using only SOHO and STEREO images resulted in a ToA $\sim$15.7 hours later than observed. Incorporating high-resolution SoloHI images significantly improved the 3D reconstruction, yielding a ToA of only $\sim$1.8 hours earlier than actual (marked by the gray vertical dashed line).} 
\label{fig:GCS_kinematics}
\end{figure*}

Figure~\ref{fig:GCS_kinematics} compares the kinematic profiles for both fits.  The CME enters the SoloHI FOV at approximately 19:00 UT, while it remains in the coronagraph fields until the leading edge exits the COR2 FOV at 20:38 UT. The GCS parameters and heights derived up until this time are very similar between the fits with and without SoloHI. Once the leading edge exits the COR2 field, LASCO/C3, and SECCHI/HI1 are the only remaining viewpoints. In C3, the event becomes too diffuse leaving the fit almost entirely dependent on just HI1. Here the two fits begin to diverge thanks to the inclusion of SoloHI data.

The enhanced spatial resolution of SoloHI relative to HI1 in a similar observing configuration was demonstrated by \citet{Hess_2023}. This alone could account for the improved fit performance with SoloHI. SoloHI runs at a higher cadence so the CME undergoes fewer changes from one image to the next, making it easier to confidently track the same feature. Finally, SoloHI provides a significantly different viewpoint compared to the LASCO/SECCHI, further reducing error. 

Neither the high-quality images from SoloHI alone nor the LASCO/SECCHI images are sufficient for an ideal input in a Space Weather forecasting framework. In the SoloHI images (top row of Figure~\ref{fig:SoloHI_STEREOHI_detail_image}), the arrival of the shock at Earth  \textit{appears\/} to be on April 22 at approximately 06:00 UT. This is clearly due to projection effects, as the interplanetary shock actually arrived at 1 AU on April 23 at 17:00 UT. A closer examination of Figure~\ref{fig:SC_orientation} reveals that SoloHI observes the flank of the CME. Meanwhile, in the LASCO and SECCHI-COR observations, the CME appears as a halo and speed estimations are highly uncertain due to projection effects. Hence, the optimal way to assess the real 3D kinematics precisely is by combining data from multiple imagers with at least one of them located away from the Sun-Earth line. SoloHI's location beyond the Sun-Earth line, combined with the high-resolution images, crucially affected the 3D reconstruction and, consequently, the prediction of the ToA at 1 au.

\section{Discussion and Conclusions} 
\label{sec:Section_5_Discussion_and_Conclusions}

The geomagnetic storm of April 23-24, 2023, was the first severe storm of Solar Cycle 25 although the CME characteristics derived from the coronagraph data (speed, mass, and kinetic energy) did not suggest such a strong impact. This event bears strong similiarities to the March 17, 2015 geomagnetic storm. The $\sim 700$ km/s CME that triggered that severe geomagnetic storm was associated with a relatively weak C9.1-class flare originating from an active region at nearly the same heliographic latitude and longitude (S22W25) as in our event. The nearly ecliptic-parallel tilt and prolonged southward magnetic field orientation of the March CME were the critical factors that led to the first severe geomagnetic storm of Solar Cycle 24 \citep[][]{Marubashi_2016_StPatricks_Storm}. 


A strong geomagnetic storm is expected when the CME magnetic field points southward (relative to Earth's dipole) for an extended period, typically on the order of several hours \citep[e.g.][]{Gonzalez_Tsurutani_1987}. The southward component of the magnetic field ($B_{z}$) reconnects with the dayside magnetopause, transferring energy from the CME into the magnetosphere \citep[e.g.][]{Gonzalez_etal_1994}. An estimate of the relationship between the minimum value of $B_{z}$, the CME speed, and the minimum value of the Dst index can be obtained from empirical relations \citep[see e.g.][]{Gonzalez_2004_Peak_Dst, Yurchyshyn_2004_SpWea}. Our single-viewpoint analysis using the COR2 images showed an asymmetric expansion, with the CME moving faster toward the south (with an average speed of 1400 km/s $\pm$ 90 km/s) compared to the north (820 km/s $\pm$ 80 km/s) with the maximum speed measured at 1520 km/s (see Section~\ref{sec:Section_4_CME_Analysis} for details). Closer to the ecliptic plane, the CME speed was approximately 1000 km/s, indicating a moderately fast CME. Equations (1) and (2) from \cite{Yurchyshyn_2004_SpWea} estimate an absolute maximum value of $B_{z}$ to be, in this case, approximately 18 nT and a minimum Dst of about -145 nT. However, the actual observed values were 34 nT and -213 nT, respectively, which far exceeds the estimates based on the 1000 km/s speed. We note at this point that these measured speeds represent lower estimates due to projection effects \citep[see e.g.,][and references therein]{Paouris_2021_ProjectionEffects}. Although these empirical relations provide rough estimates, the significant underestimation of the April severe storm highlights the need for a deeper analysis.

One of the main objectives in space weather forecasting is early and accurate prediction of the magnitude and duration of the southward component in the transient \citep[see e.g.][and references therein]{Vourlidas_etal_2019}. 
This is a formidable task, but an early estimate can be made by examining how the CME's magnetic field (most crucially, the entrained MFR) is organized at the time of the eruption.

The MFR magnetic field can be decomposed into two components: the helical (poloidal) field that wraps around the MFR and the axial (toroidal) field that runs along the central axis \citep[see e.g.][]{Chen_2017_Flux_Ropes}. The poloidal field can be twisted with left- or right-handed chirality (i.e. sign of magnetic helicity). If the chirality and orientation of the magnetic field can be estimated, then the MFR can be classified into one of eight different types according to \cite{Bothmer_&_Schwenn_1998}.

At present, there is no effective methodology to directly observe the orientation of the CME magnetic field in the corona. Various indirect proxies are used to estimate the ``intrinsic" MFR type \citep[see e.g.][]{Mostl_etal_2008, Palmerio_etal_2017, Palmerio_etal_2018}. These proxies rely on a combination of extreme ultraviolet (EUV), X-ray, and photospheric magnetogram observations to estimate the magnetic structure of CMEs either just before or at the time of eruption.  These near-Sun observations are usually combined with in situ observations at L1, to estimate the impact of CMEs in the geospace. 

For example, \cite{Palmerio_etal_2017} derived the intrinsic MFR type for two CMEs by using multi-wavelength remote sensing observations to determine the chirality of the erupting MFRs and their inclination and direction of their axial field. They found that the MFR type was consistent between the coronal estimates and the in situ L1 measurements. However, such a consistency is not reached for every CME due to various phenomena that can affect their propagation direction and orientation, potentially changing the classification type of the MFR. These phenomena include deflections \citep[][]{Kay_2013_deflections, Kay_2015_Deflection}, rotations \citep[][]{Vourlidas_et_al_2013_CMEs_Fluxropes, Isavnin_2014_rotations}, deformations \citep[][]{Isavnin_2016_FRiED} and distortions \citep[][]{Savani_etal_2010_deformations}.

\cite{Palmerio_etal_2018} extended their previous analysis with 18 additional CMEs, finding that only 4 cases out of a total of 20 (20\%) matched the MFR types observed at the Sun and in situ. 
The Sun-L1 match improved to 55\% when they considered as a match intermediate cases where the orientation at the Sun and the latitude of the in situ reconstructed MFR axis were within $35^{\circ}-55^{\circ}$. It is important to remember that the in situ data are one-dimensional, so the CME passage over the trajectory of a single spacecraft may well be unrepresentative of the global shape and orientation of the MFR. The internal structure of a CME is far from the oversimplified three-part structure model \citep[see e.g.][]{Riley_2008_3Part_CME, Chen_2017_Flux_Ropes} as evidenced by the Wide-field Imager for Solar PRobe \citep[WISPR;][]{Vourlidasetal2016_WISPR} 
onboard Parker Solar Probe \citep[Parker;][]{Foxetal2016_PSP}. These data show previously unseen features \citep[see e.g.][]{Howard_etal_2022, Paouris_etal_2024_KHI}. The MFR type derived from in situ measurements depends on where the spacecraft crosses the CME (closer to the axis), and any local distortions present within the CME may distort the results. The nature of in situ measurements requires strong assumptions about the 3D structure of the MC and raises concerns about the reliability of in situ MFR fittings for describing the large-scale structure of the MFR \citep[][]{Riley_2004_flux_rope_fitting}.

Recently, \cite{Regnault_2024_2degree_separation} showed using SolO and Wind that even angular separations as small as $2.2^{\circ}$ can significantly alter the observed CME properties. Similar differences were also reported in \cite{Palmerio_2024_PSP_Bepi} using in situ data from Parker and BepiColombo. They noted that despite the $4^{\circ}$ longitudinal and 0.03 au radial separation between the two spacecraft the in situ measurements differed notably, making it particularly challenging to understand the overall 3D CME structure.

\cite{Weiss_2024_April2023Storm} addressed the complexity of CME geometries by developing a distorted magnetic flux rope (DMFR) model to describe the structures observed during multipoint in situ measurements. Their modeling study of the April 2023 CME event relied on magnetic field measurements by Wind and STEREO, that exhibited significant differences between the two spacecraft. The MFR inclination differed by $75^{\circ}$ between the reconstructions from the individual spacecraft, underscoring the limitation of such reconstructions, particularly from single-spacecraft time series. 
The study attempted to overcome this limitation via a two-point reconstruction, assuming a very distorted MFR, which managed to reduce the inclination difference to $49^{\circ}$. However, challenges remained in aligning timing and localized features, particularly for STEREO. In any case, \cite{Weiss_2024_April2023Storm} derived the same MFR configuration (SWN) with their model as we did using the MVA approach.
The difference between \cite{Weiss_2024_April2023Storm} results and ours can be attributed to the sensitivity of both the DMFR and MVA approaches to the selection of MFR time interval boundaries. When we applied the MVA analysis using the same MFR time intervals for Wind and STEREO as in \cite{Weiss_2024_April2023Storm}, we found a significant inclination deviation of approximately $43^{\circ}$, closely matching \cite{Weiss_2024_April2023Storm} findings. Specifically, for Wind, we estimate $\theta_{MVA} \sim 12^\circ$, and for STEREO, $\theta_{MVA} \sim 55^\circ$, indicating a highly distorted MFR. This emphasizes the critical role of carefully selecting time intervals when interpreting in situ magnetic field data. The differences between Wind and STEREO, both in terms of magnetic field strength and orientation, underline the importance of considering these distortions when using in situ data to infer global CME properties. This reinforces the notion that single spacecraft crossings, as seen in earlier studies, may not provide a comprehensive view of the CME's internal structure or its evolution as it propagates through the heliosphere. 

Using multi-channel EUV images and in situ Wind data, we tracked the orientation of the CME MFR through its Sun-Earth transit. At the Sun, we estimated that the MFR was a right-handed (RH) South-West-North (SWN) with a low inclination ($|\tau| \sim 34^{\circ}$) structure. The tilt, derived from the GCS model, indicated a rotation by approximately $14^{\circ}$ within the first few hours after ejection. The MFR appeared to continue its rotation for another $4^\circ$ during the transit to Earth according to the MVA analysis at L1 ($\theta_{MVA} \sim -16^{\circ}$). Similar results were obtained with the STEREO magnetic field data ($\theta_{MVA} \sim -17^{\circ}$). Furthermore, we derived a small latitudinal deflection for the CME, as the associated solar flare was located at $-21^{\circ}$, but the GCS model resulted in a latitude of $-13^{\circ}$. The deflection of the CME towards the solar equator, is likely due to its interaction with the fast wind stream originating from the coronal hole located south of AR13283 (see the panel on the right in the first row of Figure~\ref{fig:ARIA_and_Beff}). The presence of a strong magnetic pressure gradient between the CME and the CH can cause the CME to be deflected away from the coronal hole region  \citep[see e.g.][and references therein]{Jin_2017, Cecere_2023, Kay_2013_deflections, Kay_2015_Deflection}. In our case, the CME was deflected closer to the solar equator by the coronal hole south of the active region, enhancing its space weather impact. This contrasts with the September 5, 2022, CME, which exhibited the opposite behavior—propagating southward and moving away from the solar equator. In the September 5, 2022, event, despite the presence of strong coronal signatures, the space weather impact would have been very weak if the CME had been Earth-directed, and it was indeed weak at Solar Orbiter's position.

{The MFR signatures in Wind and STEREO-A are both consistent with a large scale SWN MFR and clearly identified with visible rotations in the magnetic hodogram plots (Figure~\ref{fig:Hodograms_comb}) However, we observe two intriguing differences in the magnetic field profiles despite the relatively small longitudinal separation of $\sim 10^{\circ}$ between the two spacecraft. The ME duration was significantly longer at STEREO (28.5 hours) compared to Wind (20.7 hours) indicating CME expansion or deformation during its inner heliospheric propagation. In contrast, the sheath region was shorter at STEREO (6.4 hours) compared to Wind (8.3 hours) which further suggests kinematic variations of the transient during its propagation. The MFR duration, on the other hand, is nearly the same; 13.9 hours at Wind and 14.6 hours at STEREO. Taken together, these discrepancies reveal that while the magnetically stronger structure of the CME (the MFR) remains coherent (and likely expanding self-similarly), the surrounding ICME structure, comprised by piled-up ambient coronal and solar wind, is much more prone to distortions during the inner heliospheric propagation of the transient. A similar conclusion was reached by \citet{Kilpua_etal2013} in comparison of in situ and remote sensing CMEs in Cycles 23 and 24. Additionally, both spacecraft measured fast solar wind in the CME wake (around 550 km/s). This is likely the fast-wind stream from the coronal hole at the south of AR13283, discussed earlier. Therefore, the in situ measurements corroborate our earlier assertion on the role of the CH on the equatorward deflection of the CME. The radial (0.033~au) and angular ($10^{\circ}$) separation between Wind and STEREO places their measurements in the mesoscale region of the parameter space, as defined by \cite{Lugaz_2018_Mesoscale} (i.e., radial separation of 0.005-0.050~au and angular separation of $1^{\circ}$-$12^{\circ}$). The observed differences between the two spacecraft confirm that the CME exhibits complex internal structuring and that even small separations can lead to significant variability in measurements.
}

Finally,  the ToA hindcast emphasized once more the challenges facing the ToA forecasting of a halo CME using speed estimates from coronagraphs along the Sun-Earth line. In our case, the common assumption to use the COR2 speed estimate along the ecliptic plane (1000 km/s) led to an overestimation of the CME arrival time by 15.7 hours when applying the empirical effective acceleration model \citep[EAM;][]{Paouris_2021_EAMv3}. More accurate forecasts require tracking the CME beyond the field of view of coronagraphs, ideally from a vantage point off the Sun-Earth line, which tends to capture the two-phase kinematics of the CME \citep[][]{Paouris_Vourlidas_2022}. The SoloHI high-resolution imaging from a viewpoint at $125^{\circ}$ off the Sun-Earth line in combination with the LASCO and SECCHI imaging afforded us a robust reconstruction of the 3D CME overcoming the redundant information in the LASCO and SECCHI images due to their small longitudinal separation ($10.5^{\circ}$). The inclusion of SoloHI significantly improved the ToA hindcast (-1.8 hours), when compared to using images from SOHO and STEREO alone (15.7 hours). This improvement clearly demonstrates the importance of observations beyond the Sun-Earth line \citep[see e.g.][]{Colaninno_2013_ToA}.

Our main findings are summarized below.

\begin{itemize}
    \item The host active region, AR13283, was relatively simple in terms of magnetic complexity. The $B_{eff}$ parameter \citep[][]{GeorgoulisandRust2007} reached a rather average maximum value of ~350 G. The mass and kinetic energy of the CME place it in the top 5\% of observed CMEs but not among the extreme ones \citep[][]{Paourisetal_5Sep2022}. Therefore, neither the source region's magnetic configuration nor the CME coronal properties can explain the strong geoeffectiveness of the eruption. 
    \item The GCS and MVA analysis indicate that the CME-entrained MFR tilted towards the ecliptic plane from its original $\sim 34^{\circ}$ to $\sim -16^{\circ}$ at L1 (sections~\ref{subsec:Subsection_3_3_Flux_Rope_II} and \ref{subsec:Subsection_4_2_GCS}). The gradual alignment of the MFR's axis with the ecliptic plane, resulting from rotation and deflection, played a critical role in the CME's geoeffectiveness. 
    \item We derive an SWN MFR type with clear rotational signatures in the magnetic hodogram plots. Despite the relatively small longitudinal separation of approximately $10^{\circ}$ between the two spacecraft, notable differences are observed in the in situ measurement profiles. In particular, the ME duration was significantly longer at STEREO-A (28.5 hours) compared to Wind (20.7 hours) while the sheath region was shorter at STEREO-A (6.4 hours) compared to Wind (8.3 hours).
    \item The high-resolution imaging from SoloHI, combined with its position off the Sun-Earth line was essential for obtaining a robust 3D reconstruction of the CME compared to relying only on the observations from LASCO and SECCHI. In turn, the derived 3D kinematics significantly improved the retrospective forecast of ToA, reducing the error between the actual and hindcasted ToA to 1.8 hours. 
\end{itemize}

In summary, our analysis covered various aspects of this event, from the host active region to its arrival on Earth. This work demonstrates the importance of tracking the evolution of the CME magnetic structure through its inner heliospheric journey to understand/assess its geoeffectiveness, as well as the vital need for imaging of the heliopshere off the Sun-Earth line to provide accurate CME kinematics for space weather forecasting. Strong solar coronal signatures alone are insufficient to describe a CME's geomagnetic impact (see, for example, the September 5, 2022 CME). Rather, when combined with an analysis of the CME's evolution, these signatures can help assess geoeffectiveness. The April 2023 CME is one such case, despite its weak solar coronal signatures, the evolution of the CME in the inner heliosphere led to the occurrence of a severe geomagnetic storm, underscoring the critical role of its propagation dynamics.

\begin{acknowledgments}
We sincerely thank the anonymous referee for their valuable comments, which have significantly enhanced the clarity and interpretation of our results. We gratefully acknowledge Erika Palmerio, Andreas J. Weiss, and Teresa Nieves-Chinchilla for their useful discussions. We are also grateful to the data providers for the data used in this work, especially to STEREO, Parker, and SolO. E.P. was supported by NASA Grant 80NSSC22K0970. A.V. and G.S. are supported by 80NSSC22K1028 and 80NSSC22K0970. P.H. is supported by the Office of Naval Research. The STEREO SECCHI data are produced by a consortium of RAL (UK), NRL (USA), LMSAL (USA), GSFC (USA), MPS (Germany), CSL (Belgium), IOTA (France), and IAS (France). SOHO is a project of international cooperation between ESA and NASA. The SOHO/LASCO data used here are produced by a consortium of the Naval Research Laboratory (USA), Max-Planck-Institut fuer Aeronomie (Germany), Laboratoire d'Astronomie (France), and the University of Birmingham (UK). SolO is a space mission of international collaboration between ESA and NASA, operated by ESA. The Solar Orbiter Heliospheric Imager (SoloHI) instrument was designed, built, and is now operated by the US Naval Research Laboratory with the support of the NASA Heliophysics Division, Solar Orbiter Collaboration Office under DPR NNG09EK11I. The SDO/HMI data are provided by the Joint Science Operations Center (JSOC) Science Data Processing (SDP). 

The code for Dst estimates is archived through Zenodo at 10.5281/zenodo.8223043 \citep[][]{Paouris_2023_Dst_zndo}. The current version is available at GitHub at https://github.com/Paouris/Dst.
\end{acknowledgments}

\bibliography{bibliography}

\begin{thebibliography}{}
\expandafter\ifx\csname natexlab\endcsname\relax\def\natexlab#1{#1}\fi
\providecommand{\url}[1]{\href{#1}{#1}}
\providecommand{\dodoi}[1]{doi:~\href{http://doi.org/#1}{\nolinkurl{#1}}}
\providecommand{\doeprint}[1]{\href{http://ascl.net/#1}{\nolinkurl{http://ascl.net/#1}}}
\providecommand{\doarXiv}[1]{\href{https://arxiv.org/abs/#1}{\nolinkurl{https://arxiv.org/abs/#1}}}

\bibitem[{{Bothmer} \& {Schwenn}(1998)}]{Bothmer_&_Schwenn_1998}
{Bothmer}, V., \& {Schwenn}, R. 1998, Annales Geophysicae, 16, 1, \dodoi{10.1007/s00585-997-0001-x}

\bibitem[{{Brueckner} {et~al.}(1995){Brueckner}, {Howard}, {Koomen}, {Korendyke}, {Michels}, {Moses}, {Socker}, {Dere}, {Lamy}, {Llebaria}, {Bout}, {Schwenn}, {Simnett}, {Bedford}, \& {Eyles}}]{Brueckneretal1995}
{Brueckner}, G.~E., {Howard}, R.~A., {Koomen}, M.~J., {et~al.} 1995, \solphys, 162, 357, \dodoi{10.1007/BF00733434}

\bibitem[{{Burlaga} {et~al.}(1981){Burlaga}, {Sittler}, {Mariani}, \& {Schwenn}}]{Burlaga_1981_MC}
{Burlaga}, L., {Sittler}, E., {Mariani}, F., \& {Schwenn}, R. 1981, \jgr, 86, 6673, \dodoi{10.1029/JA086iA08p06673}

\bibitem[{{Burton} {et~al.}(1975){Burton}, {McPherron}, \& {Russell}}]{Burton_1975_Dst}
{Burton}, R.~K., {McPherron}, R.~L., \& {Russell}, C.~T. 1975, \jgr, 80, 4204, \dodoi{10.1029/JA080i031p04204}

\bibitem[{{C{\'e}cere} {et~al.}(2023){C{\'e}cere}, {Costa}, {Cremades}, \& {Stenborg}}]{Cecere_2023}
{C{\'e}cere}, M., {Costa}, A., {Cremades}, H., \& {Stenborg}, G. 2023, Frontiers in Astronomy and Space Sciences, 10, 1260432, \dodoi{10.3389/fspas.2023.1260432}

\bibitem[{{Chen}(2017)}]{Chen_2017_Flux_Ropes}
{Chen}, J. 2017, Physics of Plasmas, 24, 090501, \dodoi{10.1063/1.4993929}

\bibitem[{{Colaninno} {et~al.}(2013){Colaninno}, {Vourlidas}, \& {Wu}}]{Colaninno_2013_ToA}
{Colaninno}, R.~C., {Vourlidas}, A., \& {Wu}, C.~C. 2013, Journal of Geophysical Research (Space Physics), 118, 6866, \dodoi{10.1002/2013JA019205}

\bibitem[{{D{\'e}moulin} \& {Dasso}(2009)}]{Demoulin_Dasso_2009_MCs}
{D{\'e}moulin}, P., \& {Dasso}, S. 2009, \aap, 507, 969, \dodoi{10.1051/0004-6361/200912645}

\bibitem[{{Domingo} {et~al.}(1995){Domingo}, {Fleck}, \& {Poland}}]{Domingoetal1995}
{Domingo}, V., {Fleck}, B., \& {Poland}, A.~I. 1995, \solphys, 162, 1, \dodoi{10.1007/BF00733425}

\bibitem[{{Dunlop} {et~al.}(1995){Dunlop}, {Woodward}, \& {Farrugia}}]{Dunlop_1995_MVA}
{Dunlop}, M.~W., {Woodward}, T.~I., \& {Farrugia}, C.~J. 1995, in ESA Special Publication, Vol. 371, Proceedings of the Cluster Workshops, Data Analysis Tools and Physical Measurements and Mission-Oriented Theory, ed. K.~H. {Glassmeier}, U.~{Motschmann}, \& R.~{Schmidt}, 33

\bibitem[{{Fox} {et~al.}(2016){Fox}, {Velli}, {Bale}, {Decker}, {Driesman}, {Howard}, {Kasper}, {Kinnison}, {Kusterer}, {Lario}, {Lockwood}, {McComas}, {Raouafi}, \& {Szabo}}]{Foxetal2016_PSP}
{Fox}, N.~J., {Velli}, M.~C., {Bale}, S.~D., {et~al.} 2016, \ssr, 204, 7, \dodoi{10.1007/s11214-015-0211-6}

\bibitem[{{Galvin} {et~al.}(2008){Galvin}, {Kistler}, {Popecki}, {Farrugia}, {Simunac}, {Ellis}, {M{\"o}bius}, {Lee}, {Boehm}, {Carroll}, {Crawshaw}, {Conti}, {Demaine}, {Ellis}, {Gaidos}, {Googins}, {Granoff}, {Gustafson}, {Heirtzler}, {King}, {Knauss}, {Levasseur}, {Longworth}, {Singer}, {Turco}, {Vachon}, {Vosbury}, {Widholm}, {Blush}, {Karrer}, {Bochsler}, {Daoudi}, {Etter}, {Fischer}, {Jost}, {Opitz}, {Sigrist}, {Wurz}, {Klecker}, {Ertl}, {Seidenschwang}, {Wimmer-Schweingruber}, {Koeten}, {Thompson}, \& {Steinfeld}}]{Galvin_2008_PLASTIC}
{Galvin}, A.~B., {Kistler}, L.~M., {Popecki}, M.~A., {et~al.} 2008, \ssr, 136, 437, \dodoi{10.1007/s11214-007-9296-x}

\bibitem[{{Georgoulis}(2008)}]{Georgoulis2008}
{Georgoulis}, M.~K. 2008, \grl, 35, L06S02, \dodoi{10.1029/2007GL032040}

\bibitem[{{Georgoulis} {et~al.}(2008){Georgoulis}, {Raouafi}, \& {Henney}}]{Georgoulis_etal_2008_ARIA}
{Georgoulis}, M.~K., {Raouafi}, N.~E., \& {Henney}, C.~J. 2008, in Astronomical Society of the Pacific Conference Series, Vol. 383, Subsurface and Atmospheric Influences on Solar Activity, ed. R.~{Howe}, R.~W. {Komm}, K.~S. {Balasubramaniam}, \& G.~J.~D. {Petrie}, 107, \dodoi{10.48550/arXiv.0706.4444}

\bibitem[{{Georgoulis} \& {Rust}(2007)}]{GeorgoulisandRust2007}
{Georgoulis}, M.~K., \& {Rust}, D.~M. 2007, \apjl, 661, L109, \dodoi{10.1086/518718}

\bibitem[{{Gonzalez} {et~al.}(2004){Gonzalez}, {dal Lago}, {Cl{\'u}a de Gonzalez}, {Vieira}, \& {Tsurutani}}]{Gonzalez_2004_Peak_Dst}
{Gonzalez}, W.~D., {dal Lago}, A., {Cl{\'u}a de Gonzalez}, A.~L., {Vieira}, L.~E.~A., \& {Tsurutani}, B.~T. 2004, Journal of Atmospheric and Solar-Terrestrial Physics, 66, 161, \dodoi{10.1016/j.jastp.2003.09.006}

\bibitem[{{Gonzalez} {et~al.}(1994){Gonzalez}, {Joselyn}, {Kamide}, {Kroehl}, {Rostoker}, {Tsurutani}, \& {Vasyliunas}}]{Gonzalez_etal_1994}
{Gonzalez}, W.~D., {Joselyn}, J.~A., {Kamide}, Y., {et~al.} 1994, \jgr, 99, 5771, \dodoi{10.1029/93JA02867}

\bibitem[{{Gonzalez} \& {Tsurutani}(1987)}]{Gonzalez_Tsurutani_1987}
{Gonzalez}, W.~D., \& {Tsurutani}, B.~T. 1987, \planss, 35, 1101, \dodoi{10.1016/0032-0633(87)90015-8}

\bibitem[{{Hess} {et~al.}(2023){Hess}, {Colaninno}, {Vourlidas}, {Howard}, \& {Stenborg}}]{Hess_2023}
{Hess}, P., {Colaninno}, R.~C., {Vourlidas}, A., {Howard}, R.~A., \& {Stenborg}, G. 2023, \aap, 679, A149, \dodoi{10.1051/0004-6361/202346907}

\bibitem[{{Howard} {et~al.}(2022){Howard}, {Stenborg}, {Vourlidas}, {Gallagher}, {Linton}, {Hess}, {Rich}, \& {Liewer}}]{Howard_etal_2022}
{Howard}, R.~A., {Stenborg}, G., {Vourlidas}, A., {et~al.} 2022, \apj, 936, 43, \dodoi{10.3847/1538-4357/ac7ff5}

\bibitem[{{Howard} {et~al.}(2008){Howard}, {Moses}, {Vourlidas}, {Newmark}, {Socker}, {Plunkett}, {Korendyke}, {Cook}, {Hurley}, {Davila}, {Thompson}, {St Cyr}, {Mentzell}, {Mehalick}, {Lemen}, {Wuelser}, {Duncan}, {Tarbell}, {Wolfson}, {Moore}, {Harrison}, {Waltham}, {Lang}, {Davis}, {Eyles}, {Mapson-Menard}, {Simnett}, {Halain}, {Defise}, {Mazy}, {Rochus}, {Mercier}, {Ravet}, {Delmotte}, {Auchere}, {Delaboudiniere}, {Bothmer}, {Deutsch}, {Wang}, {Rich}, {Cooper}, {Stephens}, {Maahs}, {Baugh}, {McMullin}, \& {Carter}}]{Howard_SECCHI_2008}
{Howard}, R.~A., {Moses}, J.~D., {Vourlidas}, A., {et~al.} 2008, \ssr, 136, 67, \dodoi{10.1007/s11214-008-9341-4}

\bibitem[{{Howard} {et~al.}(2020){Howard}, {Vourlidas}, {Colaninno}, {Korendyke}, {Plunkett}, {Carter}, {Wang}, {Rich}, {Lynch}, {Thurn}, {Socker}, {Thernisien}, {Chua}, {Linton}, {Koss}, {Tun-Beltran}, {Dennison}, {Stenborg}, {McMullin}, {Hunt}, {Baugh}, {Clifford}, {Keller}, {Janesick}, {Tower}, {Grygon}, {Farkas}, {Hagood}, {Eisenhauer}, {Uhl}, {Yerushalmi}, {Smith}, {Liewer}, {Velli}, {Linker}, {Bothmer}, {Rochus}, {Halain}, {Lamy}, {Auch{\`e}re}, {Harrison}, {Rouillard}, {Patsourakos}, {St. Cyr}, {Gilbert}, {Maldonado}, {Mariano}, \& {Cerullo}}]{Howard_etal_2020_SolOHI}
{Howard}, R.~A., {Vourlidas}, A., {Colaninno}, R.~C., {et~al.} 2020, \aap, 642, A13, \dodoi{10.1051/0004-6361/201935202}

\bibitem[{{Huttunen} {et~al.}(2005){Huttunen}, {Schwenn}, {Bothmer}, \& {Koskinen}}]{Huttunen_2005_MCs}
{Huttunen}, K.~E.~J., {Schwenn}, R., {Bothmer}, V., \& {Koskinen}, H.~E.~J. 2005, Annales Geophysicae, 23, 625, \dodoi{10.5194/angeo-23-625-2005}

\bibitem[{{Isavnin}(2016)}]{Isavnin_2016_FRiED}
{Isavnin}, A. 2016, \apj, 833, 267, \dodoi{10.3847/1538-4357/833/2/267}

\bibitem[{{Isavnin} {et~al.}(2014){Isavnin}, {Vourlidas}, \& {Kilpua}}]{Isavnin_2014_rotations}
{Isavnin}, A., {Vourlidas}, A., \& {Kilpua}, E.~K.~J. 2014, \solphys, 289, 2141, \dodoi{10.1007/s11207-013-0468-4}

\bibitem[{{Janvier} {et~al.}(2013){Janvier}, {D{\'e}moulin}, \& {Dasso}}]{Janvier_2013_MCs}
{Janvier}, M., {D{\'e}moulin}, P., \& {Dasso}, S. 2013, \aap, 556, A50, \dodoi{10.1051/0004-6361/201321442}

\bibitem[{{Jarolim} {et~al.}(2024){Jarolim}, {Veronig}, {Purkhart}, {Zhang}, \& {Rempel}}]{Jarolim_2024_May2024_Superstorm}
{Jarolim}, R., {Veronig}, A., {Purkhart}, S., {Zhang}, P., \& {Rempel}, M. 2024, arXiv e-prints, arXiv:2409.08124, \dodoi{10.48550/arXiv.2409.08124}

\bibitem[{{Jin} {et~al.}(2017){Jin}, {Manchester}, {van der Holst}, {Sokolov}, {T{\'o}th}, {Vourlidas}, {de Koning}, \& {Gombosi}}]{Jin_2017}
{Jin}, M., {Manchester}, W.~B., {van der Holst}, B., {et~al.} 2017, \apj, 834, 172, \dodoi{10.3847/1538-4357/834/2/172}

\bibitem[{{Kaiser} {et~al.}(2008){Kaiser}, {Kucera}, {Davila}, {St. Cyr}, {Guhathakurta}, \& {Christian}}]{Kaiseretal2008}
{Kaiser}, M.~L., {Kucera}, T.~A., {Davila}, J.~M., {et~al.} 2008, \ssr, 136, 5, \dodoi{10.1007/s11214-007-9277-0}

\bibitem[{{Kay} {et~al.}(2013){Kay}, {Opher}, \& {Evans}}]{Kay_2013_deflections}
{Kay}, C., {Opher}, M., \& {Evans}, R.~M. 2013, \apj, 775, 5, \dodoi{10.1088/0004-637X/775/1/5}

\bibitem[{{Kay} {et~al.}(2015){Kay}, {Opher}, \& {Evans}}]{Kay_2015_Deflection}
---. 2015, \apj, 805, 168, \dodoi{10.1088/0004-637X/805/2/168}

\bibitem[{{Kay} \& {Palmerio}(2024)}]{Kay_2024_LLAMACORE}
{Kay}, C., \& {Palmerio}, E. 2024, Space Weather, 22, e2023SW003796, \dodoi{10.1029/2023SW003796}

\bibitem[{Kilpua {et~al.}(2013)Kilpua, Isavnin, Vourlidas, Koskinen, \& Rodriguez}]{Kilpua_etal2013}
Kilpua, E. K.~J., Isavnin, A., Vourlidas, A., Koskinen, H. E.~J., \& Rodriguez, L. 2013, Annales Geophysicae, 31, 1251–1265, \dodoi{10.5194/angeo-31-1251-2013}

\bibitem[{{Kontogiannis}(2024)}]{Kontogiannis_2024_May2024_Superstorm}
{Kontogiannis}, I. 2024, arXiv e-prints, arXiv:2409.18697, \dodoi{10.48550/arXiv.2409.18697}

\bibitem[{{Lemen} {et~al.}(2012){Lemen}, {Title}, {Akin}, {Boerner}, {Chou}, {Drake}, {Duncan}, {Edwards}, {Friedlaender}, {Heyman}, {Hurlburt}, {Katz}, {Kushner}, {Levay}, {Lindgren}, {Mathur}, {McFeaters}, {Mitchell}, {Rehse}, {Schrijver}, {Springer}, {Stern}, {Tarbell}, {Wuelser}, {Wolfson}, {Yanari}, {Bookbinder}, {Cheimets}, {Caldwell}, {Deluca}, {Gates}, {Golub}, {Park}, {Podgorski}, {Bush}, {Scherrer}, {Gummin}, {Smith}, {Auker}, {Jerram}, {Pool}, {Soufli}, {Windt}, {Beardsley}, {Clapp}, {Lang}, \& {Waltham}}]{Lemenetal2012_AIA}
{Lemen}, J.~R., {Title}, A.~M., {Akin}, D.~J., {et~al.} 2012, \solphys, 275, 17, \dodoi{10.1007/s11207-011-9776-8}

\bibitem[{{Lepping} {et~al.}(1995){Lepping}, {Ac{\~{u}}na}, {Burlaga}, {Farrell}, {Slavin}, {Schatten}, {Mariani}, {Ness}, {Neubauer}, {Whang}, {Byrnes}, {Kennon}, {Panetta}, {Scheifele}, \& {Worley}}]{Lepping_1995_MFI_Wind}
{Lepping}, R.~P., {Ac{\~{u}}na}, M.~H., {Burlaga}, L.~F., {et~al.} 1995, \ssr, 71, 207, \dodoi{10.1007/BF00751330}

\bibitem[{{Lin} {et~al.}(1995){Lin}, {Anderson}, {Ashford}, {Carlson}, {Curtis}, {Ergun}, {Larson}, {McFadden}, {McCarthy}, {Parks}, {R{\`e}me}, {Bosqued}, {Coutelier}, {Cotin}, {D'Uston}, {Wenzel}, {Sanderson}, {Henrion}, {Ronnet}, \& {Paschmann}}]{Lin_1995_PM_3DP_Wind}
{Lin}, R.~P., {Anderson}, K.~A., {Ashford}, S., {et~al.} 1995, \ssr, 71, 125, \dodoi{10.1007/BF00751328}

\bibitem[{{Liu} {et~al.}(2012){Liu}, {Hoeksema}, {Scherrer}, {Schou}, {Couvidat}, {Bush}, {Duvall}, {Hayashi}, {Sun}, \& {Zhao}}]{Liu_2012_HMI_vs_MDI}
{Liu}, Y., {Hoeksema}, J.~T., {Scherrer}, P.~H., {et~al.} 2012, \solphys, 279, 295, \dodoi{10.1007/s11207-012-9976-x}

\bibitem[{{Liu} {et~al.}(2024){Liu}, {Hu}, {Zhao}, {Chen}, \& {Wang}}]{Liu_2024_May2024_Superstorm}
{Liu}, Y.~D., {Hu}, H., {Zhao}, X., {Chen}, C., \& {Wang}, R. 2024, \apjl, 974, L8, \dodoi{10.3847/2041-8213/ad7ba4}

\bibitem[{{Lugaz} {et~al.}(2012){Lugaz}, {Farrugia}, {Davies}, {M{\"o}stl}, {Davis}, {Roussev}, \& {Temmer}}]{Lugaz_2012_deflection_ME}
{Lugaz}, N., {Farrugia}, C.~J., {Davies}, J.~A., {et~al.} 2012, \apj, 759, 68, \dodoi{10.1088/0004-637X/759/1/68}

\bibitem[{{Lugaz} {et~al.}(2018){Lugaz}, {Farrugia}, {Winslow}, {Al-Haddad}, {Galvin}, {Nieves-Chinchilla}, {Lee}, \& {Janvier}}]{Lugaz_2018_Mesoscale}
{Lugaz}, N., {Farrugia}, C.~J., {Winslow}, R.~M., {et~al.} 2018, \apjl, 864, L7, \dodoi{10.3847/2041-8213/aad9f4}

\bibitem[{{Luhmann} {et~al.}(2008){Luhmann}, {Curtis}, {Schroeder}, {McCauley}, {Lin}, {Larson}, {Bale}, {Sauvaud}, {Aoustin}, {Mewaldt}, {Cummings}, {Stone}, {Davis}, {Cook}, {Kecman}, {Wiedenbeck}, {von Rosenvinge}, {Acuna}, {Reichenthal}, {Shuman}, {Wortman}, {Reames}, {Mueller-Mellin}, {Kunow}, {Mason}, {Walpole}, {Korth}, {Sanderson}, {Russell}, \& {Gosling}}]{Luhmann_2008_IMPACT}
{Luhmann}, J.~G., {Curtis}, D.~W., {Schroeder}, P., {et~al.} 2008, \ssr, 136, 117, \dodoi{10.1007/s11214-007-9170-x}

\bibitem[{{Marubashi} {et~al.}(2015){Marubashi}, {Akiyama}, {Yashiro}, {Gopalswamy}, {Cho}, \& {Park}}]{Marubashi_2015_PIL_and_PEAs}
{Marubashi}, K., {Akiyama}, S., {Yashiro}, S., {et~al.} 2015, \solphys, 290, 1371, \dodoi{10.1007/s11207-015-0681-4}

\bibitem[{{Marubashi} {et~al.}(2016){Marubashi}, {Cho}, {Kim}, {Kim}, {Park}, \& {Ishibashi}}]{Marubashi_2016_StPatricks_Storm}
{Marubashi}, K., {Cho}, K.-S., {Kim}, R.-S., {et~al.} 2016, Earth, Planets and Space, 68, 173, \dodoi{10.1186/s40623-016-0551-9}

\bibitem[{{M{\"o}stl} {et~al.}(2008){M{\"o}stl}, {Miklenic}, {Farrugia}, {Temmer}, {Veronig}, {Galvin}, {Vr{\v{s}}nak}, \& {Biernat}}]{Mostl_etal_2008}
{M{\"o}stl}, C., {Miklenic}, C., {Farrugia}, C.~J., {et~al.} 2008, Annales Geophysicae, 26, 3139, \dodoi{10.5194/angeo-26-3139-2008}

\bibitem[{{M{\"u}ller} {et~al.}(2020){M{\"u}ller}, {St. Cyr}, {Zouganelis}, {Gilbert}, {Marsden}, {Nieves-Chinchilla}, {Antonucci}, {Auch{\`e}re}, {Berghmans}, {Horbury}, {Howard}, {Krucker}, {Maksimovic}, {Owen}, {Rochus}, {Rodriguez-Pacheco}, {Romoli}, {Solanki}, {Bruno}, {Carlsson}, {Fludra}, {Harra}, {Hassler}, {Livi}, {Louarn}, {Peter}, {Sch{\"u}hle}, {Teriaca}, {del Toro Iniesta}, {Wimmer-Schweingruber}, {Marsch}, {Velli}, {De Groof}, {Walsh}, \& {Williams}}]{Muller_etal_2020_Solar_Orbiter}
{M{\"u}ller}, D., {St. Cyr}, O.~C., {Zouganelis}, I., {et~al.} 2020, \aap, 642, A1, \dodoi{10.1051/0004-6361/202038467}

\bibitem[{{Nieves-Chinchilla} {et~al.}(2018){Nieves-Chinchilla}, {Vourlidas}, {Raymond}, {Linton}, {Al-haddad}, {Savani}, {Szabo}, \& {Hidalgo}}]{Nieves-Chinchilla_2018_Wind_MCs}
{Nieves-Chinchilla}, T., {Vourlidas}, A., {Raymond}, J.~C., {et~al.} 2018, \solphys, 293, 25, \dodoi{10.1007/s11207-018-1247-z}

\bibitem[{{O'Brien} \& {McPherron}(2000)}]{OBrien_McPherron_2000_Dst}
{O'Brien}, T.~P., \& {McPherron}, R.~L. 2000, \jgr, 105, 7707, \dodoi{10.1029/1998JA000437}

\bibitem[{{Ogilvie} {et~al.}(1995){Ogilvie}, {Chornay}, {Fritzenreiter}, {Hunsaker}, {Keller}, {Lobell}, {Miller}, {Scudder}, {Sittler}, {Torbert}, {Bodet}, {Needell}, {Lazarus}, {Steinberg}, {Tappan}, {Mavretic}, \& {Gergin}}]{Ogilvie_1995_SWE_Wind}
{Ogilvie}, K.~W., {Chornay}, D.~J., {Fritzenreiter}, R.~J., {et~al.} 1995, \ssr, 71, 55, \dodoi{10.1007/BF00751326}

\bibitem[{{Palmerio} {et~al.}(2017){Palmerio}, {Kilpua}, {James}, {Green}, {Pomoell}, {Isavnin}, \& {Valori}}]{Palmerio_etal_2017}
{Palmerio}, E., {Kilpua}, E.~K.~J., {James}, A.~W., {et~al.} 2017, \solphys, 292, 39, \dodoi{10.1007/s11207-017-1063-x}

\bibitem[{{Palmerio} {et~al.}(2018){Palmerio}, {Kilpua}, {M{\"o}stl}, {Bothmer}, {James}, {Green}, {Isavnin}, {Davies}, \& {Harrison}}]{Palmerio_etal_2018}
{Palmerio}, E., {Kilpua}, E.~K.~J., {M{\"o}stl}, C., {et~al.} 2018, Space Weather, 16, 442, \dodoi{10.1002/2017SW001767}

\bibitem[{{Palmerio} {et~al.}(2024){Palmerio}, {Carcaboso}, {Khoo}, {Salman}, {S{\'a}nchez-Cano}, {Lynch}, {Rivera}, {Pal}, {Nieves-Chinchilla}, {Weiss}, {Lario}, {Mieth}, {Heyner}, {Stevens}, {Romeo}, {Zhukov}, {Rodriguez}, {Lee}, {Cohen}, {Rodr{\'\i}guez-Garc{\'\i}a}, {Whittlesey}, {Dresing}, {Oleynik}, {Jebaraj}, {Fischer}, {Schmid}, {Richter}, {Auster}, {Fraschetti}, \& {Mierla}}]{Palmerio_2024_PSP_Bepi}
{Palmerio}, E., {Carcaboso}, F., {Khoo}, L.~Y., {et~al.} 2024, \apj, 963, 108, \dodoi{10.3847/1538-4357/ad1ab4}

\bibitem[{{Paouris}(2023)}]{Paouris_2023_Dst_zndo}
{Paouris}, E. 2023, {Paouris/Dst: Dst estimates}, v1.0,  Zenodo, \dodoi{10.5281/zenodo.8223043}

\bibitem[{{Paouris} {et~al.}(2024){Paouris}, {Stenborg}, {Linton}, {Vourlidas}, {Howard}, \& {Raouafi}}]{Paouris_etal_2024_KHI}
{Paouris}, E., {Stenborg}, G., {Linton}, M.~G., {et~al.} 2024, \apj, 964, 139, \dodoi{10.3847/1538-4357/ad2208}

\bibitem[{{Paouris} {et~al.}(2021{\natexlab{a}}){Paouris}, {{\v{C}}alogovi{\'c}}, {Dumbovi{\'c}}, {Mays}, {Vourlidas}, {Papaioannou}, {Anastasiadis}, \& {Balasis}}]{Paouris_2021_EAMv3}
{Paouris}, E., {{\v{C}}alogovi{\'c}}, J., {Dumbovi{\'c}}, M., {et~al.} 2021{\natexlab{a}}, \solphys, 296, 12, \dodoi{10.1007/s11207-020-01747-4}

\bibitem[{{Paouris} \& {Vourlidas}(2022)}]{Paouris_Vourlidas_2022}
{Paouris}, E., \& {Vourlidas}, A. 2022, Space Weather, 20, e2022SW003070, \dodoi{10.1029/2022SW003070}

\bibitem[{{Paouris} {et~al.}(2023){Paouris}, {Vourlidas}, {Kouloumvakos}, {Papaioannou}, {Jagarlamudi}, \& {Horbury}}]{Paourisetal_5Sep2022}
{Paouris}, E., {Vourlidas}, A., {Kouloumvakos}, A., {et~al.} 2023, \apj, 956, 58, \dodoi{10.3847/1538-4357/acf30f}

\bibitem[{{Paouris} {et~al.}(2021{\natexlab{b}}){Paouris}, {Vourlidas}, {Papaioannou}, \& {Anastasiadis}}]{Paouris_2021_ProjectionEffects}
{Paouris}, E., {Vourlidas}, A., {Papaioannou}, A., \& {Anastasiadis}, A. 2021{\natexlab{b}}, Space Weather, 19, e02617, \dodoi{10.1029/2020SW002617}

\bibitem[{{Pesnell} {et~al.}(2012){Pesnell}, {Thompson}, \& {Chamberlin}}]{Pesnelletal2012}
{Pesnell}, W.~D., {Thompson}, B.~J., \& {Chamberlin}, P.~C. 2012, \solphys, 275, 3, \dodoi{10.1007/s11207-011-9841-3}

\bibitem[{{Regnault} {et~al.}(2024){Regnault}, {Al-Haddad}, {Lugaz}, {Farrugia}, {Yu}, {Zhuang}, \& {Davies}}]{Regnault_2024_2degree_separation}
{Regnault}, F., {Al-Haddad}, N., {Lugaz}, N., {et~al.} 2024, \apj, 962, 190, \dodoi{10.3847/1538-4357/ad1883}

\bibitem[{{Riley} {et~al.}(2008){Riley}, {Lionello}, {Miki{\'c}}, \& {Linker}}]{Riley_2008_3Part_CME}
{Riley}, P., {Lionello}, R., {Miki{\'c}}, Z., \& {Linker}, J. 2008, \apj, 672, 1221, \dodoi{10.1086/523893}

\bibitem[{{Riley} {et~al.}(2004){Riley}, {Linker}, {Lionello}, {Miki{\'c}}, {Odstrcil}, {Hidalgo}, {Cid}, {Hu}, {Lepping}, {Lynch}, \& {Rees}}]{Riley_2004_flux_rope_fitting}
{Riley}, P., {Linker}, J.~A., {Lionello}, R., {et~al.} 2004, Journal of Atmospheric and Solar-Terrestrial Physics, 66, 1321, \dodoi{10.1016/j.jastp.2004.03.019}

\bibitem[{{Savani} {et~al.}(2010){Savani}, {Owens}, {Rouillard}, {Forsyth}, \& {Davies}}]{Savani_etal_2010_deformations}
{Savani}, N.~P., {Owens}, M.~J., {Rouillard}, A.~P., {Forsyth}, R.~J., \& {Davies}, J.~A. 2010, \apjl, 714, L128, \dodoi{10.1088/2041-8205/714/1/L128}

\bibitem[{{Scherrer} {et~al.}(1995){Scherrer}, {Bogart}, {Bush}, {Hoeksema}, {Kosovichev}, {Schou}, {Rosenberg}, {Springer}, {Tarbell}, {Title}, {Wolfson}, {Zayer}, \& {MDI Engineering Team}}]{Scherrer_1995_MDISOHO}
{Scherrer}, P.~H., {Bogart}, R.~S., {Bush}, R.~I., {et~al.} 1995, \solphys, 162, 129, \dodoi{10.1007/BF00733429}

\bibitem[{{Scherrer} {et~al.}(2012){Scherrer}, {Schou}, {Bush}, {Kosovichev}, {Bogart}, {Hoeksema}, {Liu}, {Duvall}, {Zhao}, {Title}, {Schrijver}, {Tarbell}, \& {Tomczyk}}]{Scherreretal2012}
{Scherrer}, P.~H., {Schou}, J., {Bush}, R.~I., {et~al.} 2012, \solphys, 275, 207, \dodoi{10.1007/s11207-011-9834-2}

\bibitem[{{Sonnerup} \& {Cahill}(1967)}]{Sonnerup_Cahill_1967_MVA}
{Sonnerup}, B.~U.~O., \& {Cahill}, L.~J., J. 1967, \jgr, 72, 171, \dodoi{10.1029/JZ072i001p00171}

\bibitem[{{Stenborg} {et~al.}(2008){Stenborg}, {Vourlidas}, \& {Howard}}]{Stenborg_2008_Wavelets}
{Stenborg}, G., {Vourlidas}, A., \& {Howard}, R.~A. 2008, \apj, 674, 1201, \dodoi{10.1086/525556}

\bibitem[{{Thernisien}(2011)}]{Thernisien_2011_GCS}
{Thernisien}, A. 2011, \apjs, 194, 33, \dodoi{10.1088/0067-0049/194/2/33}

\bibitem[{{Thernisien} {et~al.}(2009){Thernisien}, {Vourlidas}, \& {Howard}}]{Thernisien_2009_GCS_and_Errors}
{Thernisien}, A., {Vourlidas}, A., \& {Howard}, R.~A. 2009, \solphys, 256, 111, \dodoi{10.1007/s11207-009-9346-5}

\bibitem[{{Thernisien} {et~al.}(2006){Thernisien}, {Howard}, \& {Vourlidas}}]{Thernisien_2006_GCS}
{Thernisien}, A.~F.~R., {Howard}, R.~A., \& {Vourlidas}, A. 2006, \apj, 652, 763, \dodoi{10.1086/508254}

\bibitem[{{Thompson} {et~al.}(2000){Thompson}, {Cliver}, {Nitta}, {Delann{\'e}e}, \& {Delaboudini{\`e}re}}]{CME_footpoints_Thompson_2000}
{Thompson}, B.~J., {Cliver}, E.~W., {Nitta}, N., {Delann{\'e}e}, C., \& {Delaboudini{\`e}re}, J.~P. 2000, \grl, 27, 1431, \dodoi{10.1029/1999GL003668}

\bibitem[{{Tripathi} {et~al.}(2004){Tripathi}, {Bothmer}, \& {Cremades}}]{Tripathi_2004_EUV_PEAs}
{Tripathi}, D., {Bothmer}, V., \& {Cremades}, H. 2004, \aap, 422, 337, \dodoi{10.1051/0004-6361:20035815}

\bibitem[{{Verbeke} {et~al.}(2023){Verbeke}, {Mays}, {Kay}, {Riley}, {Palmerio}, {Dumbovi{\'c}}, {Mierla}, {Scolini}, {Temmer}, {Paouris}, {Balmaceda}, {Cremades}, \& {Hinterreiter}}]{Verbeke_2023}
{Verbeke}, C., {Mays}, M.~L., {Kay}, C., {et~al.} 2023, Advances in Space Research, 72, 5243, \dodoi{10.1016/j.asr.2022.08.056}

\bibitem[{{Vourlidas} {et~al.}(2010){Vourlidas}, {Howard}, {Esfandiari}, {Patsourakos}, {Yashiro}, \& {Michalek}}]{Vourlidas_etal_2010_CME_mass_energy}
{Vourlidas}, A., {Howard}, R.~A., {Esfandiari}, E., {et~al.} 2010, \apj, 722, 1522, \dodoi{10.1088/0004-637X/722/2/1522}

\bibitem[{{Vourlidas} {et~al.}(2013){Vourlidas}, {Lynch}, {Howard}, \& {Li}}]{Vourlidas_et_al_2013_CMEs_Fluxropes}
{Vourlidas}, A., {Lynch}, B.~J., {Howard}, R.~A., \& {Li}, Y. 2013, \solphys, 284, 179, \dodoi{10.1007/s11207-012-0084-8}

\bibitem[{{Vourlidas} {et~al.}(2019){Vourlidas}, {Patsourakos}, \& {Savani}}]{Vourlidas_etal_2019}
{Vourlidas}, A., {Patsourakos}, S., \& {Savani}, N.~P. 2019, Philosophical Transactions of the Royal Society of London Series A, 377, 20180096, \dodoi{10.1098/rsta.2018.0096}

\bibitem[{{Vourlidas} {et~al.}(2016){Vourlidas}, {Howard}, {Plunkett}, {Korendyke}, {Thernisien}, {Wang}, {Rich}, {Carter}, {Chua}, {Socker}, {Linton}, {Morrill}, {Lynch}, {Thurn}, {Van Duyne}, {Hagood}, {Clifford}, {Grey}, {Velli}, {Liewer}, {Hall}, {DeJong}, {Mikic}, {Rochus}, {Mazy}, {Bothmer}, \& {Rodmann}}]{Vourlidasetal2016_WISPR}
{Vourlidas}, A., {Howard}, R.~A., {Plunkett}, S.~P., {et~al.} 2016, \ssr, 204, 83, \dodoi{10.1007/s11214-014-0114-y}

\bibitem[{{Weiss} {et~al.}(2024){Weiss}, {Nieves-Chinchilla}, \& {M{\"o}stl}}]{Weiss_2024_April2023Storm}
{Weiss}, A.~J., {Nieves-Chinchilla}, T., \& {M{\"o}stl}, C. 2024, \apj, 975, 169, \dodoi{10.3847/1538-4357/ad7940}

\bibitem[{{Wu} {et~al.}(2016){Wu}, {Liou}, {Lepping}, {Hutting}, {Plunkett}, {Howard}, \& {Socker}}]{Wu_2016_StPatricks_Storm}
{Wu}, C.-C., {Liou}, K., {Lepping}, R.~P., {et~al.} 2016, Earth, Planets and Space, 68, 151, \dodoi{10.1186/s40623-016-0525-y}

\bibitem[{{Wuelser} {et~al.}(2004){Wuelser}, {Lemen}, {Tarbell}, {Wolfson}, {Cannon}, {Carpenter}, {Duncan}, {Gradwohl}, {Meyer}, {Moore}, {Navarro}, {Pearson}, {Rossi}, {Springer}, {Howard}, {Moses}, {Newmark}, {Delaboudiniere}, {Artzner}, {Auchere}, {Bougnet}, {Bouyries}, {Bridou}, {Clotaire}, {Colas}, {Delmotte}, {Jerome}, {Lamare}, {Mercier}, {Mullot}, {Ravet}, {Song}, {Bothmer}, \& {Deutsch}}]{Wuelser_STA_EUVI_2004}
{Wuelser}, J.-P., {Lemen}, J.~R., {Tarbell}, T.~D., {et~al.} 2004, in Society of Photo-Optical Instrumentation Engineers (SPIE) Conference Series, Vol. 5171, Telescopes and Instrumentation for Solar Astrophysics, ed. S.~{Fineschi} \& M.~A. {Gummin}, 111--122, \dodoi{10.1117/12.506877}

\bibitem[{{Yurchyshyn}(2008)}]{Yurchyshyn_2008_PEAs_and_flux_rope}
{Yurchyshyn}, V. 2008, \apjl, 675, L49, \dodoi{10.1086/533413}

\bibitem[{{Yurchyshyn} {et~al.}(2004){Yurchyshyn}, {Wang}, \& {Abramenko}}]{Yurchyshyn_2004_SpWea}
{Yurchyshyn}, V., {Wang}, H., \& {Abramenko}, V. 2004, Space Weather, 2, S02001, \dodoi{10.1029/2003SW000020}

\end{thebibliography}
\bibliographystyle{aasjournal}

\end{document}